\documentclass[useAMS,usenatbib,usegraphicx]{mn2e}
\usepackage{journals}
\usepackage{times}
\usepackage{color}
\title[Mass inversion of spherical systems]
{Kinematic deprojection and mass inversion of spherical systems of known
velocity anisotropy} 

\author[Gary A. Mamon
and Gwena\"el Bou\'e]{Gary A.
    Mamon$^{1,2}$\thanks{E-mail: gam@iap.fr} 
and Gwena\"el Bou\'e$^{1,3}$\thanks{E-mail: boue@imcce.fr}\\ $^1$Institut
d'Astrophysique de
    Paris (UMR 7095: CNRS \& UPMC), 98 bis Bd Arago, F--75014 Paris, France
    \\
$^2$ Astrophysics \& BIPAC, University of Oxford, Keble Rd, Oxford OX1 3RH,
UK\\
$^3$ Astronomie et Syst\`emes Dynamiques, IMCCE-CNRS UMR8028, Observatoire de
Paris, UPMC, 77 Av. Denfert-Rochereau, F-75014 Paris, France
}

\date{Accepted 2009 October 2. Received 2009 October 1; in original form 2009
  June 29}
\pubyear{2009}

\begin{document}
\onecolumn
\maketitle

\begin{abstract}
Traditionally, the degeneracy between the unknown radial profiles of total mass
and velocity 
anisotropy  inherent in the spherical,
stationary, non-streaming Jeans equation has been handled by assuming a mass
profile and fitting models to the observed kinematical data.
However, mass profiles are still not well known: there are discrepancies in
the inner slopes of the density profiles of halos found in dissipationless
cosmological $N$-body simulations, and the inclusion of gas alters
significantly the inner slopes of both the total mass and the 
dark matter component.
Here, the opposite approach is considered: 
the equation of anisotropic kinematic projection is inverted for known
arbitrary anisotropy to yield the space radial velocity dispersion profile in
terms of an integral involving the radial profiles of  anisotropy and
isotropic dynamical pressure (itself a single integral of observable quantities).
Then, through the
Jeans equation,
the mass profile of a spherical system is derived in terms of
double integrals of
observable quantities.
Single integral formulas for both deprojection and mass inversion 
are provided for several simple anisotropy models
 (isotropic,
radial, circular, general constant, Osipkov-Merritt, Mamon-{\L}okas and
Diemand-Moore-Stadel).
Tests of the mass inversion on NFW models with the first four of these anisotropy models
yield accurate results in the case of perfect
observational data, and typically better than 70\% (in 4 cases out of 5)
accurate mass profiles 
for the sampling errors expected from current observational data on clusters of galaxies.
For the NFW model with mildly increasing radial anisotropy, 
the mass is found to be insensitive to
the adopted anisotropy profile at 7 scale radii and to the adopted anisotropy
radius at 3 scale radii.
This anisotropic mass inversion method is a useful complementary tool to
analyze the mass and anisotropy profiles of spherical systems.
It provides the practical means to lift the
mass-anisotropy degeneracy in quasi-spherical systems such as 
globular clusters, round dwarf spheroidal and elliptical
galaxies, as well as groups and clusters of galaxies,
when the anisotropy of the tracer is expected to be linearly
related to the slope of its density (Hansen \& Moore 2006).
\end{abstract}

\begin{keywords}
stellar dynamics --  dark matter --  methods: analytical --  galaxies:
kinematics and dynamics --  
galaxies: haloes -- galaxies: clusters, general
\end{keywords}

\section{Introduction}
The major goal of kinematical modelling of a self-gravitating astrophysical
system, observed at one instant, is
to measure
on one hand the total mass distribution (visible and dark matter), and on the
other hand the three dimensional velocity streaming and dispersion  moments.
In other words, the modeller wishes to deduce the distributions of dark
matter and of orbital shapes.
The modeller has at his disposal, at best, maps of surface
density (or surface brightness)
and of the velocity field at each point, or else its moments
(line-of-sight mean velocity, dispersion, skewness and kurtosis). 

The basic equation for such kinematical modelling is the \emph{collisionless
Boltzmann equation} (hereafter CBE, but also often called Liouville or
Vlasov, see \citealp{Henon82}), which states the incompressibility of the
system in 6-dimensional phase (position,velocity) space:
\[
{\partial f \over\partial t} + {\bf v} \cdot
\nabla f 
- \nabla \Phi \cdot {\partial f \over \partial{\bf v}} = 0 \ ,
\]
where $\Phi$ is the gravitational potential
(hereafter potential) and $f$ is the \emph{distribution function},
that is the density in phase space.
Unfortunately, the resolution of the CBE is difficult, especially when
projection 
equations are taken into account.\footnote{Note that for \emph{dynamical}
  studies, a fast computer code
has been recently developed by \cite{AC05} that solves the CBE in 1+1
dimensions, and a 3+3 dimension code is under development by G. Lavaux and 
these authors.}
In particular, the CBE presents a degeneracy between the unknown
potential and the unknown velocity field (given that observations usually
limit the velocities to their projection along the line-of-sight, measured
through redshifts). 

The traditional simpler approach has been to use the (first) velocity moments
of the CBE, which are more easily related to observables,
the \emph{Jeans equations} that pertain to local dynamical equilibrium 
\begin{equation}
{\partial \overline{\bf v} \over \partial t}
+ (\overline{\bf v} \cdot \nabla) \overline{\bf v} = -  \nabla \Phi
- {1\over \rho}\,\nabla \cdot \left (\rho\, \mbox{\boldmath$\sigma^2$} \right )
\ ,
\label{jeans4vec}
\end{equation}
where 
$\rho$ is the space density of the \emph{tracer} used to observe the system,
$\mbox{\boldmath$\sigma^2$}$ is the tracer's dispersion tensor, whose elements
are
$\sigma_{ij}^2 = \overline{v_i\,v_j} - \overline{v_i}\,\overline{v_j}$, and
$\rho\,\mbox{\boldmath$\sigma^2$}$ is  the \emph{anisotropic dynamical
pressure tensor} of the tracer.
With the simplifying assumptions of stationarity and the absence of
streaming motions, equation~(\ref{jeans4vec}) simplifies to 
the \emph{stationary non-streaming Jeans equations}:
\begin{equation}
\nabla \cdot \left (\rho \,\mbox{\boldmath$\sigma^2$} \right )
= -\rho\,\nabla \Phi
\ .
\label{jeans2}
\end{equation}
Using the stationary non-streaming Jeans equations (\ref{jeans2}), 
one can relate the orbital properties, contained in the pressure term with
the mass distribution contained in the potential (through
Poisson's equation).

The small departures from circular symmetry of many astrophysical systems
observed in projection, such as globular clusters, the rounder 
elliptical galaxies (classes E0 to E2),
and groups and clusters of galaxies, has encouraged dynamicists to assume
spherical symmetry to perform the kinematical modelling.
The \emph{stationary non-streaming spherical Jeans equation} can then be
simply written 
\begin{equation}
{{\rm d} \left (\rho \,\sigma_r^2 \right ) \over {\rm d}r} + 2\,{\beta \over
  r}\, 
\rho \,\sigma_r^2 
= - \rho(r)\,{GM(r)\over r^2} \ ,
\label{jeanssph}
\end{equation}
where $M(r)$ is the \emph{total} mass profile, while 
\[
\beta(r) = 1 - {\sigma_\theta^2+\sigma_\phi^2\over 2\,\sigma_r^2} = 1 -
{\sigma_\theta^2 \over \sigma_r^2}
\ ,
\]
is
the tracer's
\emph{velocity anisotropy} (hereafter, \emph{anisotropy}) profile,
with $\sigma_r \equiv \sigma_{rr}$, etc., $\sigma_\theta =
\sigma_\phi$, by spherical symmetry,
and with $\beta = 1$, 0, $\to -\infty$ for radial, isotropic and circular orbits,
respectively. 
The stationary non-streaming
spherical Jeans equation provides an excellent estimate of the mass profile,
given all other 3D quantities,
in slowly-evolving triaxial systems such as halos in dissipationless
cosmological simulations 
 \citep*{TBW97} and elliptical galaxies formed by mergers of gas-rich spirals
 in dissipative $N$-body simulations
\citep{Mamon+06}.

Again, one is left with having two unknown quantities, the radial profiles of
mass and velocity anisotropy, linked by a single
equation. 
In other words, we have to deal with a serious \emph{mass-anisotropy
degeneracy}.

The simplest and most popular 
approach is to assume parametric forms for both the mass and
anisotropy profiles. One can then express the product of the observable
quantities: \emph{surface density} profile $\Sigma(R)$ and \emph{line of
  sight square velocity dispersion} profile $\sigma_{\rm los}^2(R)$  
vs. \emph{projected radius} 
$R$ 
through the  
\emph{anisotropic kinematic projection equation} \citep{BM82}
expressing the \emph{projected dynamical pressure} 
$P = \Sigma \sigma_{\rm los}^2$:
\begin{eqnarray}
P(R) = \Sigma(R)\,\sigma_{\rm los}^2(R)
&=& 2\,\int_R^\infty \left [\left(r^2-R^2\right)\,\sigma_r^2 + R^2
  \sigma_\theta^2 \right] \,\rho\,{{\rm d} r\over r\,\sqrt{r^2-R^2}} 
\label{anisprojgen}
\\
&=&2\,\int_R^\infty \left (1 - \beta\,{R^2\over r^2} \right )
p\,{r\,{\rm d}r\over \sqrt{r^2-R^2}} \ ,
\label{anisproj}
\end{eqnarray}
where equation~(\ref{anisproj}) is only valid for non-circular orbits,
and where 
$p = \rho \sigma_r^2$
is the \emph{radial dynamical pressure}.

Inserting the radial pressure\footnote{For clarity, 
we hereafter drop
  the term \emph{dynamical} before \emph{pressure}.} (eq.~[\ref{anisproj}])
in the spherical
stationary Jeans equation~(\ref{jeanssph}), one determines the line of sight
velocity dispersions essentially through a double integration over
$\rho\,M\,{\rm d}r$. 
\citeauthor{ML05b} (\citeyear{ML05b}, Appendix) have
simplified the problem by writing the projected 
pressure as a single integral
\begin{equation}
P(R) = \Sigma(R)\,\sigma_{\rm los}^2(R) 
= 2\,G\,\int_R^\infty K_{\rm proj} [r,R|\beta(r)]\,\rho\,M\,{{\rm d}r\over r}
= 2\,\int_R^\infty K_{\rm proj} [r,R|\beta(r)]\,\rho\,v_c^2 \,{\rm dr}
\ ,
\label{Pwkernel}
\end{equation}
where they were able to determine simple analytical expressions for 
the dimensionless kernel $K_{\rm proj}$ for several
popular analytical formulations of $\beta(r)$. Note that if one assumes
isotropy, the equations are greatly simplified, and one finds
\citep{Tremaine+94,PS97}  
$K_{\rm proj}(r,R)=\sqrt{1-R^2/r^2}$. 
Also, if $\beta = \hbox{cst} \neq 0$, the kernel can be
expressed either in terms of incomplete 
Beta functions \citep{ML05b}, or in terms of the
easier to compute regularized incomplete Beta functions \citep{ML06b}.
With parametric choices of the mass profile $M(r)$ and anisotropy profile
$\beta(r)$, one can fit for the free parameters of these two profiles that
lead to the best match of the observed line of sight velocity dispersion
profile. 
The drawback of this \emph{indirect method}, even with the recent
introduction of these 
simplifying kernels, is that the analysis is \emph{doubly-parametric}, so
that the 
derived parameters will be meaningless if
one does not choose the correct form for both the mass and anisotropy profiles.

The next step in complexity is to perform a \emph{single-parametric} analysis:
either isotropy is assumed to \emph{directly} obtain the mass profile, which
we call the \emph{mass inversion}, which is the focus of the present paper. 
Alternatively, a mass profile can be assumed and \emph{one
directly} determines the anisotropy profile through the 
\emph{anisotropy inversion}, first derived by \cite{BM82}, with later and
progressively simpler solutions found by \cite{Tonry83}, \cite{BBCK89},
\cite{SS90}, and \cite{DM92}.
One can attempt to lift the {mass-anisotropy degeneracy} by considering
together the variation with projected radius of the 
line-of-sight velocity dispersion \emph{and kurtosis}
\citep{Lokas02,LM03}.  
For
halos in cosmological simulations, which are not far from spherical
(\citealp{JS02} and references therein) and nearly
isotropic (\citealp{ML05b} and references therein), viewed in projection, 
this dispersion-kurtosis
analysis yields fairly accurate masses, concentrations and anisotropies
\citep*{SLM04}.
Unfortunately,
the line-of-sight projection of the 4th order Jeans equation, required in the
dispersion-kurtosis method, is only possible
when $\beta = \rm cst$, which does not appear to be realistic for elliptical
galaxies formed by major mergers \citep{Dekel+05}.

An even more sophisticated and general approach is to adopt a potential and
minimize the residuals between the predicted
and true observables, i.e. the distribution of objects in projected phase space
$(R,v_{\rm los})$ (where $v_{\rm los}$ is the line of sight velocity) 
by one of several methods involving the distribution
function:
\vspace{-0.5\baselineskip}
\begin{enumerate}
\item A general global form for the distribution function is adopted, 
in terms of known integrals of motions. For example, in spherical systems
with isotropic non-streaming
velocities, the distribution function is a function of energy
only, while in anisotropic non-streaming spherical systems it is a 
function of energy and the modulus of the angular momentum.
Alas,
there is no known realistic form for $f=f(E,{\bf J})$ for anisotropic
non-streaming spherical systems nor for
non-spherical systems, although \cite{Wojtak+08} have recently shown that
cosmological halos have distribution functions that can be written
$f(E,J) =
f_E(E)\,J^{2\,(\beta_\infty-\beta_0)}\,\left(1+J^2/J_0^2\right)^{-\beta_0}$,
where we adopt hereafter the notations $\beta_0 = \beta(0)$ and $\beta_\infty
= \lim_{r\to\infty} \beta$, where $J_0$ is a free parameter related to the
`anisotropy' radius where $\beta(r) = (\beta_0+\beta_\infty)/2$. Unfortunately,
\citeauthor{Wojtak+08} do not provide an analytical formula for $f_E(E)$.
\item A set of elementary distribution functions of $E$ or $(E,J)$ is chosen,
as first proposed by \cite{Dejonghe89}, then \cite{MS93}, and applied to
elliptical galaxies by \cite{GJSB98}. One then searches the linear
combination of these distribution functions, with positive weights (to ensure
a positive global distribution function) that minimizes the residuals between
the predicted and true observables.
However, there is no guarantee that the set of elementary distribution
functions constitute a basis set, so that some global realistic distribution
functions may be missed. Moreover, the distribution function may depend on an
additional unknown integral of motion.
\item A set of delta- distribution functions, $f = f(E,{\bf J})$ is chosen,
  in other words one 
considers 
orbits of given $E$ and $\bf J$ \citep{Schwarzschild79,RT84,ST96}. Again
one searches for a linear combination of these orbits that minimises the
residuals between predicted and true observables, again enforcing positive
weights. These weights are obtained either by
averaging the observables over an orbit (\citeauthor{Schwarzschild79}) or by 
continuously updating them (\citeauthor{ST96}; \citealp{dLDGS07}).
This method is powerful enough to handle non-spherical potentials.
Despite concerns about convergence 
\citep{CE04,VME04}, the orbit-superposition method, if properly implemented,
does reproduce the correct solutions \citep{Richstone+04,Thomas+04}.
\end{enumerate}
\vspace{-0.5\baselineskip}
%

The potential can be adapted from the observations, assuming constant
mass-to-light ratio ($M/L$) if the observed density is a surface brightness,
or constant mass-to-number ratio ($M/N$)
if the observed density is a surface number density.
If spherical symmetry is assumed, this involves
a choice of $M/L$ or $M/N$, 
the deprojection of the surface density map, and
then Poisson's
equation is easily inverted to obtain the potential from the density.
For axisymmetric systems, one can deproject
the surface density maps into a potential assuming it to be
the sum of gaussians \citep*{EMB94}.
One can add to the potential a possible dark component given in
parametric form (see, e.g., \citealp{WBC09}).


%

Alternatively, instead of using distribution functions, one can fit the
distribution of objects in projected phase space by 
the multiple parametric adjustment of the mass and anisotropy profiles, as well as
possibly the velocity distribution in space (which could be non-gaussian, see
\citealp{KMM04,WLGM05,HMZS06}), as in the MAMPOSSt method (Mamon, Biviano \&
Bou\'e, in preparation). 

Returning to direct single-parametric estimations, 
the mass profile of astronomical systems does not seem to be
better established than the anisotropy profile.
Indeed, despite early claims (\citealp*{NFW96}, hereafter NFW) 
of a universal density profile for
the structures (\emph{halos}) in dissipationless cosmological $N$-body
simulations of a flat Universe of cold dark matter with a cosmological constant
(hereafter $\Lambda$CDM), 
there has been an ongoing debate on whether the inner slope is
steeper \citep{FM97,Moore+99} or shallower \citep{Navarro+04,SWTS02,Stoehr06}.
Furthermore, the inclusion of gas in cosmological simulations can lead to much
steeper dark matter density
profiles \citep{GKKN04}. 
Indeed, the dissipative nature of baryons leads them to accumulate
in the 
centers of systems, not only in spiral galaxies, as is well known, but also 
in elliptical galaxies, for otherwise the NFW-like mass distribution as found 
in $\Lambda$CDM halos would lead to a local $M/L$ and aperture velocity
dispersion much lower than observed
\citep{ML05a}, and the dominance of baryons in the center and dark matter in
the envelopes has been recently confirmed by 
X-ray measurements \citep{Humphrey+06}. Moreover, the
dark matter dynamically  
responds to the baryons that dominate in the inner regions, to reach steeper
slopes than they would have had without the presence of baryons
(\citealp{BFFP86,GKKN04}). But the final density profile of dark matter is
expected to be very sensitive to the details of the baryonic feedback
processes.

On the other hand, the anisotropy profiles of the halos
in dissipationless cosmological
simulations appears to be fairly universal (see the compilation by
\citealp{ML05b} and references therein, and \citealp{Wojtak+08}), although
galaxy-mass halos have somewhat more radial orbits than cluster-sized halos
\citep{AG08}.
Also a similar shape of anisotropy 
profile holds in 
$N$-body+SPH simulations of merging spirals galaxies, including gas, but with a
ratio of anisotropy to virial radius that is ten times smaller \citep{Dekel+05}.
Moreover, dissipationless $N$-body simulations (cosmological and binary
mergers) indicate that the 
anisotropy is linearly related to the slope of the density profile
\citep{HM06}, although the trend is less clear in elliptical galaxies
formed in $N$-body+SPH simulations as merger remnants of spiral galaxies
\citep{Mamon+06}, because of the dynamical interaction of the stellar, dark
matter and dissipative gas components.

In this paper, we derive and test the mathematics of the mass inversion.
We begin in Sect.~\ref{deprojiso} with a reminder on the kinematic deprojection
of isotropic systems, followed by the mass inversion of isotropic systems in
Sect.~\ref{massiso}.
We then develop in Sect.~\ref{deprojanis}
our algorithm for the kinematic deprojection of anisotropic
systems, and in Sect.~\ref{massanis} we deduce the
mass profile with the Jeans equation~(\ref{jeanssph}).  
In Sect.~\ref{tests}, we test our mass inversion methods.

The reader in a hurry might want to
skip the mathematical details. (S)he will find the
general anisotropic deprojection formulae in equation~(\ref{qgen4}), with
special cases given in
equations
(\ref{prad1}) [radial orbits],
(\ref{pthetasol2}) [circular orbits],
and
in equations~(\ref{qgenspec}) [constant $\beta < 1$, Osipkov-Merritt,
  Mamon-{\L}okas, and Diemand-Moore-Stadel], 
with $C_\beta$ given in Table~\ref{shortcuts} and  
kernels $K_{\rm i}$ given in 
equations~(\ref{Kcst}) [constant $\beta < 1$],
(\ref{KOM}) [Osipkov-Merritt],
(\ref{KML}) [Mamon-{\L}okas],
and
(\ref{KDMS}) [Diemand-Moore-Stadel].
The formulae for the mass inversion will be found in
equations~(\ref{M2}) [general], 
(\ref{vcrad2}) [radial], 
(\ref{vccirc}) [circular], and
(\ref{vcgenspec}) [constant $\beta < 1$, Osipkov-Merritt, Mamon-{\L}okas, and
  Diemand-Moore-Stadel] with the same $C_\beta$ and kernels, and with
$D_\beta$ also given in Table~\ref{shortcuts}.

In the very late stages of this work, we came across a draft of
\cite{Wolf+09}, who independently developed an analogous method for
anisotropic kinematic deprojection. While \citeauthor{Wolf+09} produce
a general formula for kinematic deprojection, the present article also
provides simpler formulae for the kinematic deprojection with specific simple
anisotropy profiles, as well as general and specific formulae for the mass
profile.

\section{Method}

\subsection{Kinematic deprojection of isotropic systems}
\label{deprojiso}

We begin by reviewing the mathematical formalism for the kinematic
deprojection of isotropic systems.
The \emph{structural projection equation}, relating the space
density 
$\rho(r)$ to the (projected) surface density $\Sigma(R)$:
\begin{equation}
\Sigma(R)
= \int_{-\infty}^\infty \rho(r) \,{\rm d}z =
2\,\int_R^\infty {\rho(r) \,r\,{\rm d}r \over \left (r^2-R^2 \right)^{1/2}}
\ ,
\label{abelproj}
\end{equation}
is inverted through the usual Abel transform,
whose derivation we recall in appendix~\ref{deprojisoapp}, as we will use it
in the following subsection. 
One then recovers 
the well-known \emph{structural deprojection} or \emph{Abel
  inversion} equation 
\begin{equation}
\rho(r)
= -{1\over \pi} \int_r^\infty
{{\rm d}\Sigma \over {\rm d}R}\,{{\rm d}R\over \left (R^2-r^2\right )^{1/2}} \ .
\label{rhoder}
\end{equation}

In the case of isotropic velocities one can express
the {projected dynamical pressure} 
$\Sigma\,\sigma_{\rm los}^2$ in terms of the {dynamical
  pressure} 
$\rho\,\sigma^2$ with the \emph{isotropic kinematical projection
  equation}, obtained by setting $\beta=0$ in the anisotropic kinematic
projection equation~(\ref{anisproj}): 
\begin{equation}
\Sigma\,\sigma_{\rm los}^2 = 2\,\int_R^\infty \rho\,\sigma^2 {r\,{\rm d}r\over
  \sqrt{r^2-R^2}} \ .
\label{isoprojec}
\end{equation}
Equation~(\ref{isoprojec}) is the strict analog to equation~(\ref{abelproj}),
where the tracer density$\rho$ is replaced by the dynamical pressure
$p = \rho\,\sigma^2$ and the surface density $\Sigma$ is replaced by
projected pressure
$P = \Sigma\,\sigma_{\rm los}^2$.\footnote{Given the isotropy, the space
velocity dispersion is equal to the radial velocity dispersion, so we drop
the subscript `$r$'.}
With these replacements, the structural 
deprojection equation~(\ref{rhoder}) turns into  
the \emph{isotropic kinematical deprojection equation}
\begin{equation}
p_{\rm iso}(r) \equiv [\rho(r)\,\sigma^2 (r)]_{\beta=0}
= - {1\over \pi}\,\int_r^\infty {{\rm d} 
P
\over {\rm d}R}\,{{\rm d}R \over \sqrt{R^2-r^2}} \ .
\label{pisodef}
\end{equation}

\subsection{Mass inversion of isotropic systems}
\label{massiso}

Now, from the stationary non-streaming spherical Jeans
equation~(\ref{jeanssph}), with the isotropic condition ($\beta=0$), the
total mass profile is trivially
\begin{equation}
M(r) = -{r^2 \over G\,\rho}\,{{\rm d}p_{\rm iso}\over {\rm d}r}
=  {1\over \pi} {r^2\over G\,\rho} \,
{{\rm d}\over {\rm d}r}\int_r^\infty {{\rm d}P\over {\rm d}R} \, {{\rm d}R \over \sqrt{R^2-r^2}}
 \ .
\label{Mtmpiso}
\end{equation}
With the variable substitution $R = r\,u$, we can avoid the singularity in
the surface term of the derivative of the integral of
equation~(\ref{pisodef}) or (\ref{Mtmpiso}) by
writing
\begin{equation}
p'_{\rm iso}(r) \equiv {{\rm d}p_{\rm iso}\over {\rm d}r} = -{1\over \pi}\,
{{\rm d}\over {\rm d}r} \int_r^\infty {{\rm d}P\over {\rm d}R} \, {{\rm d}R \over \sqrt{R^2-r^2}}
= -{1\over \pi}\,\int_1^\infty P''(ru) {u\,{\rm d}u\over \sqrt{u^2-1}} = -{1\over \pi\,r}\,\int_r^\infty
{{\rm d}^2P\over{\rm d}R^2}\, {R\,{\rm d}R\over \sqrt{R^2-r^2}} \ ,
\label{pprimeiso}
\end{equation}
where $P''(R) = {\rm d}^2P/{\rm d}R^2$.
Inserting the right-hand-side of equation~(\ref{pprimeiso}) into the first
equality of
equation~(\ref{Mtmpiso}),  
we then obtain the \emph{isotropic mass inversion equation}  
\begin{equation}
M(r) 
= - {r\over G} \,{\displaystyle \int_r^\infty 
{{\rm d}^2 \left (\Sigma \sigma_{\rm los}^2 \right )\over {\rm d}R^2} 
\, {R\,{\rm d}R \over \sqrt{R^2-r^2}}
\over \displaystyle \int_r^\infty 
{{\rm d}\Sigma\over {\rm d}R}\,{{\rm d}R\over \sqrt{R^2-r^2}}} \ ,
\label{Miso}
\end{equation}
where we used the structural deprojection equation~(\ref{rhoder}) to replace
the density in the denominator.
The isotropic mass inversion equation can be further simplified, expressing the
\emph{circular velocity}, $v_c^2 = GM/r$ as\footnote{We've
  never encountered in the literature the mass profile written in this direct
  fashion, although \cite{Romanowsky+09} gave the equivalent expression
  \[
v_c^2(r) = {r^2\over \pi\rho(r)}\,\int_r^\infty {{\rm d} (P'/R)\over {\rm
    d}R}\,{{\rm d}R\over \sqrt{R^2-r^2}} \ .
\] The expression in equation~(\ref{vciso}) seems
  preferable as the differentiation is performed in a single pass.}
\begin{equation}
v_c^2(r) 
= {1\over \pi \rho(r)}\,\int_r^\infty {{\rm d}^2P\over {\rm d}R^2}\,{R\,{\rm
    d}R\over \sqrt{R^2-r^2}} \ .
\label{vciso}
\end{equation}
Unfortunately, the mass and circular velocity profiles require the second
derivative of the (observable)
projected pressure $P = \Sigma \sigma_{\rm los}^2
(R)$. The singularity 
$\left (R^2-r^2\right)^{-1/2}$ in the numerators of equations~(\ref{Miso}) and
(\ref{vciso}) prevents one from expressing the mass profile with single
integrals derivatives of the projected dynamical pressure after a suitable
integration by parts.

\subsection{Kinematic deprojection of systems of arbitrary known anisotropy
  profile} 
\label{deprojanis}

\subsubsection{General anisotropy}

The anisotropic kinematic projection equation~(\ref{anisproj})
is strictly valid for non-circular orbits (finite
$\beta$). 
For circular orbits ($\sigma_r=0$)
equation~(\ref{anisprojgen}) yields 
\begin{equation}
P(R)= 2\,R^2\int_R^\infty p_\theta {{\rm d}r\over r\,\sqrt{r^2-R^2}} \ ,
\label{Pcirc}
\end{equation}
where 
\[
p_\theta = (1-\beta)\,p=\rho \,\sigma_\theta^2
\] 
is the \emph{tangential dynamical pressure}.
To guide the reader, 
Table~\ref{nomen} reviews the nomenclature adopted in this paper.
\begin{table}
\caption{Nomenclature\label{nomen}}
\begin{tabular}{lll}
\hline
Definition & full expression & abbreviated expression \\
\hline
space radius & & $r$\\
projected radius & & $R$ \\
projected pressure & $\Sigma \,\sigma_{\rm los}^2$ & $P$ \\
radial pressure & $\rho\,\sigma_r^2$ & $p$ \\
tangential pressure &  $\rho\,\sigma_\theta^2$ & $p_\theta = (1-\beta)\,p$ \\
anisotropy & $1-\sigma_\theta^2/\sigma_r^2$ & $\beta$ \\
circular velocity & $\sqrt{GM(r)/r}$ & $v_c$ \\
\hline
\end{tabular} 
\end{table}

We repeat the steps of the standard (isotropic) Abel inversion
(Appendix~\ref{deprojisoapp}), now defining 
\begin{eqnarray}
J(r) &=& \int_r^\infty P \,{R\,{\rm d}R\over \sqrt{R^2-r^2}}
\label{Jdef}\\
&=& -\int_r^\infty {{\rm d}P \over
{\rm d}R}\,\sqrt{R^2-r^2}\,{\rm d}R \ ,
\label{Jparts}
\end{eqnarray}
where  equation~(\ref{Jparts}) is obtained by integration by parts
(the surface term is 0 for $P(R) \propto R^{-\alpha}$ with $\alpha > 1$).

For non-circular orbits, inserting the projected pressure
(eq.~[\ref{anisproj}]) into the definition of $J$ (eq.~[\ref{Jdef}]), one finds
\begin{eqnarray}
J(r) &=& 2\,\int_r^\infty {R\,{\rm d}R\over \sqrt{R^2-r^2}}\,\int_R^\infty
\left (1 - \beta\,{R^2\over s^2} \right )
p\,{s\,{\rm d}s\over \sqrt{s^2-R^2}} \ , \label{J1} \\
&=& 2\,\int_r^\infty \!\!\! p\,s\,{\rm d}s\,\int_r^s \!\!
{R\,{\rm d}R\over \sqrt{\left (R^2-r^2\right)\,\left(s^2-R^2\right)}}
- 2\,\int_r^\infty \!\!\! \beta\, p\,{{\rm d}s\over s}\,\int_r^s
\!\!  {R^3\,{\rm d}R\over  \sqrt{\left (R^2-r^2\right)\,\left(s^2-R^2\right)}} \ ,
\label{J2}  \\
&=& {\pi\over 2}\,\int_r^\infty \left [2-\left ({r^2\over
  s^2}+1\right)\,\beta\right] p\,s\,{\rm d}s \ ,
\label{J3}
\end{eqnarray}
where equation~(\ref{J2}) is obtained after reversing the order of
integration and the two inner integrals of eq.~(\ref{J2}) are worth $\pi/2$
and $(\pi/4) (r^2+s^2)$, respectively.
Differentiating equation~(\ref{J3}), one has
\begin{equation}
{{\rm d}J \over {\rm d}r} = -\pi\,r\,\left [(1-\beta)\,p + \int_r^\infty
  \beta \,p \,{{\rm d}s\over s} \right ] \ .
\label{dJdr2}
\end{equation}
Now, equation~(\ref{Jparts}) can be differentiated to yield
\begin{equation}
{{\rm d}J\over {\rm d}{\rm r}} = r\,\int_r^\infty {{\rm d}P \over
{\rm d}R}\,{{\rm d}R\over \sqrt{R^2-r^2}} 
= -\pi\,r\,p_{\rm iso}(r) \ ,
\label{dJdr3} 
\end{equation}
where the second equality in equation~(\ref{dJdr3}) comes from
equation~(\ref{pisodef}).
Equations~(\ref{dJdr2}) and (\ref{dJdr3}) yield 
\begin{equation}
p_\theta(r) 
= p_{\rm iso}(r)
- \int_r^\infty \beta\, p\,{{\rm d}s\over s} \ .
\label{impliciteqp}
\end{equation}







Equation~(\ref{impliciteqp}) is an implicit integral equation for $p$ 
with $p_{\rm iso}$ (eq.~[\ref{pisodef}]) and
$\beta$ known. 
For finite $\beta < 1$, we solve for $p$ by
differentiating 
equation~(\ref{impliciteqp}), to get the differential equation
\begin{equation}
p' - {r\,\beta' + \beta \over 1 - \beta }\,{p\over r} = {p'_{\rm iso}\over
  1-\beta} \ .
\label{pprime}
\end{equation}
Now, if we write
\begin{equation}
p' - {r\,\beta' + \beta \over 1 - \beta }\,{p\over r} 
= {1\over g}\,{{\rm d} (g\,p)\over {\rm d}r} \ ,
\label{pprimewg}
\end{equation}
then equations~(\ref{pprime}) and (\ref{pprimewg}) lead to
\begin{equation}
p(r) = -{1\over g(r)}\,\int_r^\infty {g\,p'_{\rm iso}\over 1-\beta}\,{\rm d}s
\ ,
\label{p1}
\end{equation}
where the upper limit at infinity ensures that the radial pressure
$p = \rho\,\sigma_r^2$
does not reach negative values at a finite radial distance.
But equation~(\ref{pprimewg}) directly gives
\[
{{\rm d}\ln f\over {\rm d}\ln r} = - {-r\,\beta' + \beta \over 1-\beta} \ ,
\]
hence
\begin{equation}
g(r) = g(r_1) \,\exp \left (-\int_{r_1}^r {s\,\beta'+\beta \over
  1-\beta}\,{{\rm d}s\over s} \right )  
\label{f}
\end{equation}
for any arbitrary $r_1$.
With equation~(\ref{f}), equation~(\ref{p1}) leads to 
\begin{eqnarray}
p(r) 
&=& -\exp\left (\int_{r_1}^r {s\,\beta'+\beta\over 1-\beta}\,{{\rm d}s\over
  s} \right ) 
\,\int_r^\infty \exp \left (-\int_{r_1}^s {t\,\beta'+\beta\over
  1-\beta}\,{{\rm d}t\over t} 
\right )\,{p'_{\rm iso}\over 1-\beta}\,{\rm d}s \ , \nonumber \\
&=& 
- \int_r^\infty \exp \left (-\int_r^s {t\,\beta'+\beta\over 1-\beta}\,{{\rm
    d}t\over t} 
\right )\,{p'_{\rm iso}\over 1-\beta}\,{\rm d}s \ ,
\label{pgen1}
\end{eqnarray}
where the second equality is obtained adopting $r_1=r$.
One wishes to avoid the second derivative of the observables that occurs in
the expression of equation~(\ref{pprimeiso}) for $p'_{\rm iso}(r)$, which will
amplify any uncertainties on the measurements of these observables.
Integrating by parts  the integral in
equation~(\ref{pgen1}), we finally obtain
%
\begin{eqnarray}
p(r) 
&=& {p_{\rm iso}(r)\over 1-\beta(r)} - \int_r^\infty p_{\rm iso}(s)
{A_\beta(r,s)\over 1-\beta(s)}\,{{\rm d}s\over s} \ ,
\label{pgen4}
\end{eqnarray}
where $p_{\rm iso}$
is given in
equation~(\ref{pisodef}), and where
\begin{equation}
A_\beta(r,s) = {\beta(s)\over
1-\beta(s)}\,
\exp\left(-\int_r^s {t\,\beta'+\beta\over 1-\beta}\,{{\rm d}t \over t} \right
) \ ,
\label{Abeta}
\end{equation}
which is provided in Table~\ref{shortcuts} for various simple anisotropy models.

\begin{table}
\begin{center}
\caption{Terms in equations~(\ref{qgen5}), (\ref{qgenspec}) and (\ref{Dbeta})
  for specific 
  anisotropy profiles\label{shortcuts}}
\tabcolsep 2mm
\begin{tabular}{lcccc}
\hline
\hline
Anisotropy model& $\displaystyle A_\beta(r,s)$ & $\displaystyle B_\beta(r,s)$ 
& $\displaystyle C_\beta(r)$ 
& $\displaystyle  D_\beta(r)$ \smallskip \\
& (eq.~[\ref{Abeta}]) & (eq.~[\ref{Bbeta}]) & (eq.~[\ref{qgenspec}]) &  (eq.~[\ref{Dbeta}]) \\
\hline
$\beta = \rm cst$ 
& $\displaystyle
{\beta\over 1-\beta}\,\left ({r\over s}\right)^{\beta/(1-\beta)}$
& $\displaystyle {\beta\over 1-\beta} 
\,\left ({r\over  s}\right )^{\beta/(1-\beta)}$ 
& $\displaystyle {1\over 2}\,{\beta\over 1-\beta}$
& $\displaystyle{(3-2\beta)\,\beta\over
  1-\beta}$  \\
\\
Osipkov-Merritt (eq.~[\ref{betaOM}]) 
& $\displaystyle \left ({s\over a}\right)^2 {r^2+a^2\over s^2+a^2}\,\exp\left
({r^2-s^2\over 2\, a^2}\right )$
& $\displaystyle \left ({s\over a}\right)^2 \exp\left
({r^2-s^2\over 2\, a^2}\right )$
& $\displaystyle {r \over a}$
& $\displaystyle \left ({r\over a}\right)^2\,{r^2 +  5\,a^2\over r^2 + a^2}$ \\
\\
Mamon-{\L}okas (eq.~[\ref{betaML}]) 
& $\displaystyle {r+a\over s+a}\,\left ({s\over s+2a}\right)$
& $\displaystyle {(r+2\,a)\,s\over (s+2\,a)^2}$
& $\displaystyle \left ({r\over a}\right)\,{r+2\,a\over a}$
& $\displaystyle{2\,r\over a+r}$ \\
\\
Diemand-Moore-Stadel (eq.~[\ref{betaDMS}]) 
& $\displaystyle s^{1/3} {\left (a^{1/3}-s^{1/3}\right)^3\over
\left (a^{1/3}-r^{1/3}\right)^4 }$ 
& $\displaystyle s^{1/3} {\left (a^{1/3}-s^{1/3}\right)^2\over
\left (a^{1/3}-r^{1/3}\right)^3 }$ 
& $\displaystyle {r\over \left (a^{1/3}-r^{1/3}\right)^3}$
& $\displaystyle {2\over 3}\left ({r\over a}\right)^{1/3} {5\,a^{1/3} -
    3\,r^{1/3}\over a^{1/3} - r^{1/3}}$ 
\\
\hline
\end{tabular} 
\end{center}

{Notes: the Diemand-Moore-Stadel values are restricted to $r<a$.}

\end{table}

One may prefer to use the tangential dynamical
  pressure instead of the radial one, as it can be expressed in 
a slightly simpler form:
\begin{eqnarray}
p_\theta(r) 
&=& 
- \int_r^\infty \exp \left (-\int_r^s {\beta\over 1-\beta}\,{{\rm d}t\over t}
\right )\,p'_{\rm iso}\,{\rm d}s \ ,
\label{qgen3}\\
&=& p_{\rm iso}(r)
- \int_r^\infty p_{\rm iso} {\beta\over
1-\beta}\,
\exp\left(-\int_r^s {\beta\over 1-\beta}\,{{\rm d}t \over t} \right
){{\rm d}s\over s} \ ,
\nonumber \\
&=& p_{\rm iso}(r) - \int_r^\infty p_{\rm iso}(s)\,B_\beta(r,s)\,{{\rm
    d}s\over s} \ ,
\label{qgen4}
\end{eqnarray}
as similarly derived in Appendix~\ref{appq}, and where 
\begin{equation}
B_{\rm \beta}(r,s) = {\beta(s)\over
1-\beta(s)}\,
\exp\left(-\int_r^s {\beta\over 1-\beta}\,{{\rm d}t \over t} \right
) \ ,
\label{Bbeta}
\end{equation}
which 
is provided again in Table~\ref{shortcuts} for our simple anisotropy models. 
The radial pressure is then
simply
$p(r) = p_\theta(r)/[1 - \beta(r)]$.

The expressions for the dynamical pressure (radial or tangential) are made of
single integrals involving $p_{\rm iso}$, which is a single integral
itself. Hence, the dynamical pressure is expressed in terms of double
integrals.
For simple anisotropy profiles, we can simplify the dynamical pressure to
single integrals by inserting the expression for $p_{\rm iso}(s)$
(eq.~[\ref{pisodef}]) in equation~(\ref{qgen4}) and inverting the order of
integration. This yields
\begin{eqnarray}
p_\theta(r) &=& p_{\rm iso}(r) + {1\over \pi}\,\int_r^\infty {{\rm d}P\over
  {\rm d}R}\,{\rm d}R
\,\int_r^R {\beta \over 1-\beta}\,\exp\left (-\int_r^s {\beta\over
  1-\beta}\,{{\rm d}t\over t}\right)\,{{\rm d}s\over s\,\sqrt{R^2-s^2}}
\nonumber \\
&=& p_{\rm iso}(r) + {1\over \pi}\,\int_r^\infty P'(R) {\rm d}R\,\int _r^R
B_\beta(r,s)\,{{\rm d}s\over s\,\sqrt{R^2-s^2}}
\label{qgen5}
\end{eqnarray}
and for simple $\beta(r)$, the inner integral can be expressed in closed
form, as we shall now see. 
  
\subsubsection{Case of finite $\beta$ = cst $<1$}

Equation~(\ref{pgen4}) with $A_\beta$ from Table~\ref{shortcuts} leads to
\begin{equation}
p(r) = 
{p_{\rm iso}(r) \over 1-\beta(r)}
- {\beta\over (1-\beta)^2}
\,r^{\beta/(1-\beta)}\,\int_r^\infty  
p_{\rm iso}
\,s^{-\beta/(1-\beta)}\,{{\rm d}s\over s}\ .
\label{piso1}
\end{equation}
Using equation~(\ref{qgen5}) with $B_\beta$ from Table~\ref{shortcuts}, one
obtains 
a single integral representation
for
the tangential dynamical pressure
\begin{eqnarray}
p_\theta(r) &=&  p_{\rm iso} (r) + {1\over \pi}\,{\beta\over
  1-\beta}\,r^{\beta/(1-\beta)}\, 
\int_r^\infty {{\rm d}P\over
  {\rm d}R}\,{\rm d}R
\,\int_r^R s^{-\beta/(1-\beta)}\,{{\rm d}s\over s\,\sqrt{R^2-s^2}} \nonumber
  \\
&=&p_{\rm iso} (r) + {1\over
    2\,\pi}\,{\beta\over 1-\beta}\,{1\over r}\,
\int_r^\infty {{\rm d}P\over
  {\rm d}R}\,K_{\rm cst} \left ({r\over R} \right )\,{\rm d}R \ ,
\label{qbetacst}
\end{eqnarray}
where the second equality of equation~(\ref{qbetacst}) 
is obtained with the change of variable $t = 1 -
s^2/R^2$.
The dimensionless kernel in equation~(\ref{qbetacst}) is
\begin{equation}
K_{\rm cst}(u) = u^{1/(1-\beta)}\,B \left (1-u^2,{1\over2},-{\beta/2\over
  1-\beta}\right ) \ ,
\label{Kcst}
\end{equation}
where 
$B(x,a,b) = \int_0^x t^{a-1}\,(1-t)^{b-1} {\rm d}t$ is the incomplete Beta
function.
Integrating by parts the integral in equation~(\ref{qbetacst}),
we finally obtain after some algebra a single integral expression for the
tangential pressure that does not depend on derivatives of the observations:
\begin{equation}
p_\theta(r) =
p_{\rm iso}(r)
+ {1\over 2\,\pi}\,{\beta\over \left (1\!-\!\beta\right)^2}
\left [
r^{\beta/(1\!-\!\beta)}
\!\!\int_r^\infty\!\!\!\!P(R)\,R^{-(2\!-\!\beta)/(1\!-\!\beta)}\,B\left
(1\!-\!{r^2\over  R^2},{1\over2},-{\beta/2\over 1\!-\!\beta}\right )\,{\rm
  d}R  
-2 (1\!-\!\beta) 
\!\!\int_r^\infty \!\!\!\!{P(R)\over \sqrt{R^2-r^2}}\,{{\rm d}R\over R}
\right ] .
\label{piso3}
\end{equation}
The surface term $R^{-1/(1\!-\!\beta)}\,B \left [1-r^2/R^2,{1/2},-\beta/(2\,(
1\!-\!\beta))\right ] P(R)$ that occurs in the integration by parts
goes to 0 as $R\to \infty$.
Indeed,
for $x=r/R$ and $c=-\beta/2/(1-\beta)$,
one has
$
x^{1-2c}\,B(1-x^2,{1/2},c) =
- {x/c} +O (x^3)
$
and moreover $P\to0$.
In practice, if a programming language does not provide the incomplete Beta
  function, but only the regularized incomplete Beta function,
 as $I(x,a,b) = B(x,a,b)/B(a,b) = \Gamma(a+b) B(x,a,b)/[\Gamma(a)\Gamma(b)]$,
one should then
be careful that
  $\Gamma(b)$ diverges when the
  last term $b$ in the incomplete Beta function is a negative integer,
  i.e. when 
$\beta = 2n/(2n+1) = 2/3, 4/5, 6/7 ...$ ($n$ being a positive integer).
Luckily, $B(x,a,b)$ always converges to finite values.\footnote{An SM macro
  for $B(x,a,b)$ is available upon request.}

In the limit $\beta \to 0$ everywhere, equations~(\ref{piso1}),
(\ref{qbetacst}), and (\ref{piso3}) all reduce to
$p(r) = p_{\rm iso}(r)$, as expected.

\subsubsection{Case of radial orbits: $\beta = 1$}
For radial orbits, differentiation of equation~(\ref{impliciteqp}) leads to
\begin{equation}
p(r)  = -r\,p'_{\rm iso} (r)
= {1\over \pi}\,\int_r^\infty  {{\rm d}^2 P\over
{\rm d}R^2}\,{R\,{\rm d}R\over \sqrt{R^2-r^2}} 
\ .
\label{prad1}
\end{equation}

\subsubsection{Case of circular orbits: $\beta \to -\infty$}

For circular orbits, 
we proceed in a similar fashion:
inserting the projected pressure
(eq.~[\ref{Pcirc}]) into the definition of $J$ (eq.~[\ref{Jdef}]), one finds
\begin{eqnarray}
J(r) &=& 2\,\int_r^\infty {R^3\,{\rm d}R\over \sqrt{R^2-r^2}}\,\int_R^\infty
p_\theta {{\rm d}s\over s\,\sqrt{s^2-R^2}} \ , \nonumber  \\
&=& {\pi\over 2}\,\int_r^\infty p_\theta \left (r^2+s^2\right )\,{{\rm d}s\over s}
\ ,
\end{eqnarray}
and
\begin{equation}
{{\rm d}J\over {\rm d}r} = -\pi \,r\,\left (p_\theta - \int_r^\infty p_\theta\,{{\rm d}s\over s} \right ) \ .
\label{dJdrcirc}
\end{equation}
Equations~(\ref{dJdr3}) and (\ref{dJdrcirc}) lead to
\begin{equation}
-{{\rm d}\over {\rm d}r}
 \left ({1\over \pi r}\,{{\rm d}J\over {\rm d}r} \right )
=
p'_\theta + {p_\theta\over r} = p'_{\rm iso}
= {1\over f}\,{{\rm d} \left (f \,p_\theta \right )\over {\rm d}r} \ ,
\end{equation}
whose solution is given by $f=r$:
\begin{equation}
p_{\theta}(r) = -{1\over r}\,\int_r^\infty p'_{\rm iso}\, s\,{\rm
    d}s 
= 
p_{\rm iso}(r) + {1\over
  r}\,\int_r^\infty p_{\rm iso}\,{\rm d}s \ ,
\label{pthetasol}
\end{equation}
where the 2nd equality is found by integration by parts, for which the
surface term, $\lim_{r\to\infty} r\,p_{\rm iso}(r)$, 
vanishes for ${\rm d}\ln \rho/{\rm d} \ln r + {\rm d}\ln M/{\rm
  d}\ln r < 0$ (as derived from the Jeans equation~[\ref{jeanssph}]), as is
the case for reasonable mass and tracer density profiles.
Inserting $p_{\rm iso}$ (eq.~[\ref{pisodef}]) into equation~(\ref{pthetasol})
and inverting the order of integration, we finally obtain the single integral
expression for the tangential pressure:
\begin{equation}
p_\theta(r) =
-{1\over \pi}\,\int_r^\infty 
{{\rm d}P\over {\rm d}R}\,
\left [ {1\over \sqrt{R^2-r^2}}+{1\over r}\,\cos^{-1} \left ({r\over
    R}\right) \right] 
\,{\rm d}R \ .
\label{pthetasol2}
\end{equation}

\subsubsection{Case of Osipkov-Merritt anisotropy}

For the Osipkov-Merritt \citep{Osipkov79,Merritt85} anisotropy
\begin{equation}
\beta(r) = {r^2\over r^2+a^2} \ ,
\label{betaOM}
\end{equation}
equations~(\ref{qgen3}) reduces to
\begin{equation}
p_\theta(r)
= - \int_r^\infty \exp \left
(-{s^2-r^2\over 
2\,a^2}\right )\,p'_{\rm iso}\,{\rm d}s 
= \left (\rho\,\sigma_r^2 \right)_{\rm iso}(r)
-{1\over a^2}\,\int_r^\infty \exp \left ({r^2-s^2\over 2\,a^2}\right
)\,p_{\rm iso}\,s\,{\rm d}s \ ,
\label{qOM1}
\end{equation}
where the last equality is again obtained after integration by parts or from
equation~(\ref{qgen4}).
Equation~(\ref{qgen5}) yields  (see Table~\ref{shortcuts})
a single integral representation for the
tangential dynamical pressure:
\begin{eqnarray}
p_\theta(r) &=& p_{\rm iso}(r) + {1\over\pi a^2}\,\exp \left ({r^2\over 2
  a^2}\right)
\,\int_r^\infty 
{{\rm d}P\over {\rm d}R}\,{\rm d}R 
\,\int_r^R \exp\left (-{s^2\over 2 a^2} \right)\,{s\,{\rm d}s\over
  \sqrt{R^2-s^2}} 
\nonumber \\
&=& 
p_{\rm iso}(r) 
+ {1\over \pi \,a}\,
\int_r^\infty {{\rm d}P\over {\rm d}R}\,K_{\rm OM} \left (\sqrt{R^2-r^2\over
  2 \,a^2}\right )\,{\rm d}R \ ,
\label{qOM2}
\end{eqnarray}
 where the dimensionless kernel $K_{\rm OM}$ is
\begin{equation}
K_{\rm OM}(u) = \sqrt{2}\,F(u) = \sqrt{\pi\over
  2}\,\exp\left(-u^2\right)\,{\rm erfi}\,u \ ,
\label{KOM}
\end{equation}
where
\[
F(u) = {\sqrt{\pi}\over 2}\,\exp \left (-u^2 \right )
{\rm erfi}\,u = {\sqrt{\pi}\,\exp \left(-u^2\right)\, {\rm erf} (iu)\over 2\,i} 
\]
is Dawson's integral and where ${\rm erfi}(x)$ is the imaginary error function.
Note that Dawson's integral
is available in most software packages for mathematical
functions.\footnote{SM macros for ${\rm erfi}(x)$ and Dawson's $F(x)$ are 
  available upon request.} 
Equation~(\ref{qOM2}) can also be found by inserting the expression
for $p_{\rm 
  iso}(s)$ (eq.~[\ref{pisodef}]) into  equation~(\ref{qOM1}) and reversing
  the order of integration.


\subsubsection{Case of Mamon-{\L}okas anisotropy}

For the simple anisotropy profile that \cite{ML05b} found to fit well
$\Lambda$CDM halos
\begin{equation}
\beta(r) = {1\over 2}\,{r\over r+a}
\ ,
\label{betaML}
\end{equation}
one obtains
\begin{equation}
p_\theta(r)
= - \left (r+2\,a\right )\,
\int_r^\infty  p'_{\rm iso}\,{{\rm d}s\over s+2\,a} 
= p_{\rm iso}(r)
-(r+2\,a)\,\int_r^\infty p_{\rm iso}\,{{\rm d}s\over (s+2\,a)^2} \ ,
\label{qML1}
\end{equation}
where the first equality is from equation~(\ref{qgen3}), while the second one
is obtained after integration by parts or  from equation~(\ref{qgen4}).
Equation~(\ref{qgen5}) now yields  (see Table~\ref{shortcuts})
the
single integral expression for the tangential dynamical pressure:
\begin{eqnarray}
p_\theta(r) = {r/2+a\over r+a}\,\rho(r)\,\sigma_r^2(r)  
&=& p_{\rm iso}(r)
+ {1\over \pi}\,(r+2\,a)\,\int_r^\infty 
{{\rm d}P\over {\rm d}R}\,
\int_r^R {{\rm d}s \over (s+2\,a)^2\,\sqrt{R^2-s^2}}
\nonumber \\
&=& p_{\rm iso}(r)
+ {1\over \pi}\,{r+2\,a\over a^2}\,\int_r^\infty 
{{\rm d}P\over {\rm d}R}\,K_{\rm ML}
\left ({R\over a},{r\over a}\right) \,
{\rm d}R \ ,
\label{qML}
\end{eqnarray}
where the dimensionless kernel $K_{\rm ML}$, using
$X=R/a$, $x=r/a$ and $y=s/a$, is
\begin{eqnarray}
K_{\rm ML}(X,x) &=& \int_x^X {{\rm d} y\over \sqrt{X^2-y^2}\, (y+2)^2} 
\nonumber \\
 &=& {1\over X^2}\,
\int_0^{\cos^{-1}(x/X)}{{\rm d}\theta\over \left (\cos \theta + 2/X
  \right)^2} 
\label{KMLeq}\\
&=& 
\left \{
\begin{array}{ll}
\displaystyle
-{1\over 4-X^2}\,{\sqrt{X^2-x^2}\over 2+x}+
{4\over \left (4-X^2\right)^{3/2}} \tan^{-1} \left [ \sqrt{2-X\over
    2+X}\,\sqrt{X-x\over X+x} \right ]
& \qquad \hbox{for } X<2 \ ,\\
\displaystyle
{ 1\over  12}\,{ (4+x)\,\sqrt{2-x}\over 
  (2+x)^{3/2}}
& \qquad \hbox{for } X=2 \ ,\\
\displaystyle
{1\over X^2-4}\,{\sqrt{X^2-x^2}\over 2+x}-
{4\over \left (X^2-4\right)^{3/2}} \tanh^{-1} \left [ \sqrt{X-2\over
    X+2}\,\sqrt{X-x\over X+x} \right ]
& \qquad \hbox{for } X>2 \ ,\\
\end{array}
\right.
\label{KML}
\end{eqnarray}
where equation~(\ref{KMLeq}) is found through the variable substitution $y
= X\, \cos \theta$. 
Equations~(\ref{qML}) and (\ref{KMLeq}) can also be found
by inserting the expression for $p_{\rm iso}$ (eq.~[\ref{pisodef}]) into
equation~(\ref{qML1}) 
and reversing the order of integration.


\subsubsection{Case of generalized Mamon-{\L}okas anisotropy}

The velocity anisotropies in halos in cosmological $N$ body simulations do
not always fit the Mamon-{\L}okas  formula (eq.~[\ref{betaML}]), but instead,
$\beta(r)$ shows halo to halo variations in its limits at $r=0$ and
$r\to\infty$ \citep{Wojtak+08}. Hence,
a more general form for the anisotropy profile is (e.g. \citealp{Tiret+07})
\begin{equation}
\beta(r) = \beta_0 + \left (\beta_\infty-\beta_0 \right )\,{r\over r+a} \ .
\label{betalin}
\end{equation}
The Mamon-{\L}okas anisotropy is the special case with $\beta_0=0$ and
$\beta_\infty=1/2$. 
For $\beta_0 < 1$ and $\beta_\infty < 1$, inserting equation~(\ref{betalin}) into
equation~(\ref{qgen3}) yields, after some algebra:
\begin{eqnarray} 
p_\theta(r) &=& -r^{\beta_0/(1-\beta_0)}\,\left [\left (1-\beta_\infty\right)\,r + \left
  (1-\beta_0 \right)\,a
  \right]^{\beta_\infty/(1-\beta_\infty)-\beta_0/(1-\beta_0)}
\nonumber \\
&\mbox{}&\qquad \times\int_r^\infty 
s^{-\beta_0/(1-\beta_0)}\,\left [\left (1-\beta_\infty\right)\,s + \left
  (1-\beta_0 \right)\,a
  \right]^{\beta_0/(1-\beta_0)-\beta_\infty/(1-\beta_\infty)} 
\,p'_{\rm iso}\,{\rm d}s \ .
\label{qbetalin}
\end{eqnarray}
For $\beta_0 < \beta_\infty = 1$, 
the same procedure gives
\begin{equation}
p_\theta(r) = - \exp \left ({r/a\over 1-\beta_0}\right )\,r^{\beta_0/(1-\beta_0)}\,
\int_r^\infty \exp \left (-{s/a\over 1-\beta_0}\right
)\,s^{-\beta_0/(1-\beta_0)}\,p'_{\rm iso}\,{\rm d}s  \ .
\label{qbetainf1}
\end{equation}
For $\beta_\infty < \beta_0 = 1$,\footnote{Decreasing anisotropy profiles are
  found for some regular halos \citep{Wojtak+08}, although the central
  anisotropy is never unity.}
we similarly obtain
\begin{equation}
p_\theta(r) = - \exp\left (-{a/r\over 1-\beta_\infty}\right)
\,r^{\beta_\infty/(1-\beta_\infty)}
\,\int_r^\infty \exp\left ({a/s\over 1-\beta_\infty}\right)
\,s^{-\beta_\infty/(1-\beta_\infty)}\,p'_{\rm iso}\,{\rm d}s \ .
\label{qbeta01}
\end{equation}
The integrals of equations~(\ref{qbetalin}),  (\ref{qbetainf1}), and
(\ref{qbeta01}), are essentially double integrals, since they involve
$p'_{\rm iso}$ (eq.~[\ref{pprimeiso}]). Single integral solutions do not
appear to be possible to 
derive, even for the simple case of $\beta_0=0$ (unless
$\beta_\infty = 1/2$, i.e. the Mamon-{\L}okas anisotropy model).


\subsubsection{Case of Diemand-Moore-Stadel anisotropy}

Finally for the other simple anisotropy profile that \citeauthor*{DMS04_vel}
(\citeyear{DMS04_vel}, Sect. 3.3.2) also
found to fit well 
$\Lambda$CDM halos
\begin{equation}
\beta(r) = \left \{
\begin{array}{cl}
\displaystyle
\vspace{2mm}
\left ({r\over a}\right)^{1/3} & r < a \ ,\\
1 & r\geq a \ , 
\end{array}
\right. 
\label{betaDMS}
\end{equation}
we obtain
\begin{equation}
\begin{array}{lll}
\displaystyle
p_\theta(r) &
 \displaystyle
= p_{\rm iso}(r) - {1\over \left (a^{1/3}-r^{1/3} \right)^3}\,
\int_r^a p_{\rm iso} \left
(a^{1/3}-s^{1/3}\right)^2\,
{{\rm d}s\over s^{2/3}} & \qquad \hbox{for } r<a \
,
\label{qDMSsmallr}
\\
\displaystyle
p(r) &
\displaystyle
= p_{\rm rad}(r) & 
\qquad  \hbox{for } r\geq a \ ,\\
\label{pDMSbigr}
\end{array}
\end{equation}
where equation~(\ref{qDMSsmallr}) is obtained from
equation~(\ref{qgen4}), while the equation~(\ref{pDMSbigr}) 
comes from the pure radial orbits for $r\geq a$
(eq.~[\ref{betaDMS}]). 
Again, for $r<a$, the integral in 
equation~(\ref{qDMSsmallr})  is essentially a double integral (because of $p'_{\rm
  iso}$), and a single integral solution can be obtained using
equation~(\ref{qgen5}), yielding (with Table~\ref{shortcuts}) 
\begin{eqnarray}
p_\theta(r) &=& 
p_{\rm iso}(r) + {1\over \pi \left (a^{1/3}-r^{1/3}\right)^3}\,\int_r^a {{\rm
    d}P\over 
  {\rm d}R}\,{\rm d}R
\,\int_r^R {\left (a^{1/3}-s^{1/3}\right)^2\over s^{2/3}\,\sqrt{R^2-s^2}}
\,{\rm d}s 
\nonumber \\
&=&
p_{\rm iso}(r) + {1\over \pi \left (a^{1/3}-r^{1/3} \right)^3}\,\int_r^a
{{\rm d}P\over 
  {\rm d}R}\,K_{\rm DMS} \left ({R\over a},{r\over a}\right)\,{\rm d}R \ ,
\label{qDMSsmallr2}
\end{eqnarray}
where the dimensionless kernel is
\begin{eqnarray}
K_{\rm DMS}(X,x) &=\!\!&
\int_x^X {\left (1-y^{1/3}\right)^2\over \sqrt{X^2-y^2}}\,{{\rm d}y\over
  y^{2/3}} 
  \nonumber \\
 &\!\!=\!\!& 
{1\over X^{2/3}}\,\int_0^{\cos^{-1}(x/X)} {{\rm d}\theta\over \cos^{2/3}\theta}
- 
{2\over X^{1/3}}\,\int_0^{\cos^{-1}(x/X)} {{\rm d}\theta\over \cos^{1/3}\theta}
+ \int_0^{\cos^{-1}(x/X)} {\rm d}\theta
\nonumber \\
 &\!\!=\!\!& 
\left [\sqrt{\pi}\,{\Gamma(1/6)\over \Gamma(2/3)}-B \left ({x^2\over
    X^2},{1\over 6},{1\over 2} \right ) \right ] \,{X^{-2/3}\over 2}
- \left [\sqrt{\pi}\,{\Gamma(1/3)\over \Gamma(5/6)}-B \left ({x^2\over
    X^2},{1\over 3},{1\over 2} \right ) \right ] \,X^{-1/3}
+ \cos^{-1} \left ({x\over X}\right ) \ ,
\label{KDMS}
\end{eqnarray}
for 
$X=R/a$, $x=r/a$ and $y=s/a$.

\subsubsection{General expression for the tangential pressure for specific
  anisotropy profiles}

The expressions for the tangential pressure for the cases of constant,
Osipkov-Merritt, Mamon-{\L}okas, and Diemand-Moore-Stadel anisotropy
(eqs.~[\ref{qbetacst}], [\ref{qOM2}], [\ref{qML}], and [\ref{qDMSsmallr2}],
respectively) can all be written in the form
\begin{equation}
p_\theta(r) = p_{\rm iso}(r) + {1\over \pi\,r}\,C_\beta(r)\,\int_r^\infty {{\rm
    d}P\over {\rm d}R}\,K_\beta\,{\rm d}R 
= {1\over \pi\,r}\,\int_r^\infty {{\rm
    d}P\over {\rm d}R}\,\left [C_\beta(r)\,K_\beta\left({R\over a},{r\over a}\right) - {r\over
    \sqrt{R^2-r^2}} 
  \right ]\,{\rm d}R \ ,
\label{qgenspec}
\end{equation}
where the second equality of equation~(\ref{qgenspec}) is found with
equation~(\ref{pisodef}) and where $C_\beta(r)$ and $K_\beta(X,x)$ are
dimensionless functions such that
\begin{equation}
C_\beta(r)\,K_\beta\left({R\over a},{r\over a}\right) = r\,\int_r^\infty
{B_\beta(r,s)\over  \sqrt{r^2-s^2}}\,{{\rm d}y\over y} \ ,
\end{equation}
with 
$C_\beta(r)$  given in Table~\ref{shortcuts}, and
$K_\beta$ given in
equations~(\ref{Kcst}), (\ref{KOM}), (\ref{KML}) and (\ref{KDMS}),
respectively.
For the Diemand-Moore-Stadel anisotropy model,
the upper integration limits in equation~(\ref{qgenspec})  should be replaced
by the anisotropy radius $a$.
The second equality of equation~(\ref{qgenspec}) allows the kinematic
deprojection with a unique single integral. 

\subsection{Mass profiles of spherical systems with arbitrary known anisotropy}
\label{massanis}

\subsubsection{General mass profile}

The mass profile is obtained through stationary non-streaming spherical 
Jeans equation~(\ref{jeanssph}), which writes
\begin{equation}
- \rho {G M\over r^2} = p' + {2\over r}\,\beta\,p
\ .
\label{jeansp}
\end{equation}
Now,
equation~(\ref{pprime}) reads
\begin{equation}
p' = 
{p'_{\rm iso}\over 1-\beta} + {r\,\beta'+\beta\over 
  1-\beta}\,{p\over r}
= {p'_{\rm iso}\over 1-\beta} + {r\,\beta'+\beta\over \left
  (1-\beta\right)^2}\,{p_\theta\over r} \ .
\label{prime}
\end{equation}
   Inserting $p'$ from equation~(\ref{prime}) into equation~(\ref{jeansp})
yields the \emph{general mass inversion equation}  (dropping the dependencies 
on $r$ for clarity):
\begin{equation}
-(1-\beta)\,\rho\,{G M\over r^2} (r) = 
p'_{\rm iso}(r) + \left [{\beta' + (3-2\,\beta)\,\beta/r
    \over 1-\beta}\right ]\,p_\theta (r)
= p'_{\rm iso}(r) + {D_\beta(r)\over r}\,p_\theta(r)
\ ,
\label{genMofr}
\end{equation}
where the dimensionless function
\begin{equation}
D_\beta(r) = {r\,{\rm d}\beta/{\rm d}r + (3-2\,\beta)\,\beta \over 1-\beta}
\label{Dbeta}
\end{equation}
is given in Table~\ref{shortcuts} for four anisotropy models.
Inserting the general expression for $p_\theta$ into
equation~(\ref{genMofr}), and converting the mass into the circular velocity
with $v_c^2(r) =G\,M(r)/r$
gives  
either 
\begin{equation}
[1-\beta(r)] \,\rho(r)\, v_c^2(r) =
 - r\,p'_{\rm iso}(r)
+ \left [{r\,\beta' + (3-2\,\beta)\,\beta \over 1-\beta}
\right ]\,
\left [\int_r^\infty p_{\rm iso}\,{\beta\over 1-\beta}
\,\exp\left (-\int_r^s {\beta\over 1-\beta}\,{{\rm d}t\over t} \right
)\,{{\rm d}s\over s}
- p_{\rm iso} \right ]
\ ,
\label{M1}
\end{equation}
(from eq.~[\ref{pgen4}])
or
\begin{equation}
[1-\beta(r)] \,\rho(r)\, v_c^2(r) =
\left [{r\,\beta' + (3-2\,\beta)\,\beta \over 1-\beta}
\right ]\,
\int_r^\infty \exp \left (-\int_r^s {\beta\over 1-\beta}\,{{\rm d}t\over t}
\right )\,p'_{\rm iso}\,{\rm d}s
- r\,p'_{\rm iso}(r)
\label{M2}
\end{equation}
(from eq.~[\ref{pgen1}]).
Alas, both forms (eqs.~[\ref{M1}] and [\ref{M2}]) involve the second derivative
of the observable $P$, hence the second form (eq.~[\ref{M2}]) seems
preferable to use as it is simpler.
However, for simple anisotropy profiles, the double integral of
equations~(\ref{M1}) 
and (\ref{M2}) can be simplified to single integrals, or equivalently, single
integral expressions for $p_\theta$ exist, which can be inserted into
equation~(\ref{genMofr}) to obtain a single integral expression for the mass
profile.  

\subsubsection{Case of isotropic systems}
For isotropic systems ($\beta=0$), equation~(\ref{genMofr}) trivially leads
to 
\begin{equation}
v_c^2(r) = - {r p'_{\rm iso}(r)\over \rho(r)}
\ ,
\end{equation}
 which is equivalent to the first equality of equation~(\ref{Mtmpiso}).

\subsubsection{Case of finite $\beta = {\rm cst} < 1$}

For finite $\beta = \rm cst < 1$, while equation~(\ref{M2})
becomes  (with $D_\beta$ from Table~\ref{shortcuts})
\begin{equation}
-\left (1-\beta\right)\,\rho\,{G M\over r^2} (r)
= p'_{\rm iso} - \left [{\beta\,(3-2\,\beta)/r \over 1-\beta}
\right ]\,
r^{\beta/(1-\beta)}\,\int_r^\infty s^{-\beta/(1-\beta)}\,p'_{\rm iso}\,{\rm d}s \ ,
\end{equation}
a single integral expression is found inserting the tangential pressure
(eq.~[\ref{piso3}]) into 
equation~(\ref{genMofr}) to yield
\begin{eqnarray}
-\left (1\!-\!\beta\right)\rho\,{G M\over r^2} (r)
&\!\!=\!\!& p'_{\rm iso} + \left [{\beta\,(3-2\,\beta)/r \over 1-\beta}
\right ]\,\left \{ p_{\rm iso} + {1\over 2\pi}\,{\beta\over
  (1-\beta)^2}\,\right.\nonumber \\
&\mbox{}& \left.
\qquad \times \left [
r^{\beta/(1\!-\!\beta)}
\!\!\int_r^\infty\!\!\!\!P(R)\,R^{-(2\!-\!\beta)/(1\!-\!\beta)}\,B\left (1\!-\!{r^2\over  R^2},{1\over2},-{\beta/2\over 1\!-\!\beta}\right )\,{\rm d}R 
-2 (1\!-\!\beta) 
\!\!\int_r^\infty \!\!\!\!{P(R)\over \sqrt{R^2-r^2}}\,{{\rm d}R\over R}
\right ] \right \} \ .
\end{eqnarray} 


\subsubsection{Case of radial orbits: $\beta=1$}

For radial anisotropy, 
equations~(\ref{prad1}) and (\ref{jeansp}) simply yield
\[
\rho(r) {G M(r) \over r^2} = 3\,p'_{\rm iso} +
r\,p''_{\rm iso} \ .
\]
However, using the change of variables $R = r\cosh u$, the last equality of 
equation~(\ref{prad1}) yields 
\begin{equation}
p'_{\rm rad} = {p_{\rm rad}\over r} + {1\over \pi r}\,\int_r^\infty P'''
R^2\,{{\rm d}R\over \sqrt{R^2-r^2}}
\ ,
\end{equation}
hence, from equation~(\ref{jeansp}):
\begin{equation}
v_c^2(r) = -{1\over \pi \,\rho(r)}\,
\int_r^\infty \left ( 3\,P'' + R\, P''' \right ) {R\,{\rm d}R\over
  \sqrt{R^2-r^2}} \ .
\label{vcrad2}
\end{equation}
Alas, equation~(\ref{vcrad2}) involves a triple differentiation of the
observables.

\subsubsection{Case of circular orbits: $\beta\to-\infty$}

For circular orbits, the first term in the 
stationary non-streaming spherical Jeans
equation~(\ref{jeanssph}) vanishes, and one is left with the trivial relation
\begin{equation}
v_c^2(r) = 
2\,\sigma_\theta^2 = {2\over \rho(r)}\,\left [p_{\rm iso} +
{1\over r}\,\int_r^\infty p_{\rm iso}\,{\rm d}s \right ] \ ,
\end{equation}
where we made use of equation~(\ref{pthetasol}) for the last equality.
Integrating the last equation by parts, or equivalently, using
equation~(\ref{pthetasol2}), we get  
\begin{equation}
v_c^2(r) = -{2\over \pi\,\rho(r)}
\,\int_r^\infty 
{{\rm d}P\over {\rm d}R}\,
\left [ {1\over \sqrt{R^2-r^2}}+{1\over r}\,\cos^{-1} \left ({r\over R}\right
    ) \right] 
\,
{\rm d}R \ .
\label{vccirc}
\end{equation}

\subsubsection{Case of Osipkov-Merritt anisotropy}

For Osipkov-Merritt anisotropy (eq.~[\ref{betaOM}]), equation~(\ref{genMofr})
leads to  (with $D_\beta$ from Table~\ref{shortcuts})
\begin{equation}
\rho\,{G M\over r^2} (r)
= -{r^2+a^2\over a^2}\,p'_{\rm iso} +
{1\over a^2}\,\left ({r^2+a^2\over a^2}+4 \right )\,r\,
\int_r^\infty \exp  \left ({r^2-s^2\over 2\,a^2}\right
)\,p'_{\rm iso}\,{\rm d}s \ .
\end{equation}
A single integral solution to the mass profile is obtained by inserting
$p_\theta$ of equation~(\ref{qOM2}) into equation~(\ref{genMofr}), yielding

\begin{equation}
-\rho\,{G M\over r^2} (r)
= {r^2+a^2\over a^2}\,p'_{\rm iso} + \left ({r^2 + 5\,a^2} \right ) \,{r\over
  a^5}\,\left [{p_{\rm
    iso}\,a} + {1\over \sqrt{2\pi}}\,
\exp \left
({r^2 \over 2 a^2}\right )\,
\int_r^\infty {{\rm d}P\over {\rm d}R}\, \exp \left (-{R^2\over 2 a^2}\right
)\,{\rm erfi} 
\sqrt{R^2-r^2\over 2 a^2}\,{\rm d}R \right ]\ .
\end{equation}

\subsubsection{Case of Mamon-{\L}okas anisotropy}

For \citeauthor{ML05b} anisotropy (eq.~[\ref{betaML}]), equation~(\ref{genMofr})
brings   (with $D_\beta$ from Table~\ref{shortcuts})
\begin{equation}
\rho\,{G M\over r^2} (r)
= - 2\,{r+a\over r+2\,a}\,p'_{\rm iso} 
+4\,
\int_r^\infty p'_{\rm iso}\,{{\rm d}s\over s+2\,a} \ .
\end{equation}
The single integral solution, found by inserting $p_\theta$ from
equation~(\ref{qML}) into equation~(\ref{genMofr}), is
\begin{equation}
-\rho\,{G M\over r^2} (r)
= {2\,\over r+2\,a}\,\left [(r+a)\,p'_{\rm iso} + 2\,p_{\rm iso}\right] + {4\over
  \pi\,a^2}\,\int_r^\infty  
{{\rm d}P\over {\rm d}R}\,K_{\rm {ML}} \left ({R\over a},{r\over a} \right
    )\,{\rm d}R \ ,
\end{equation}
where the dimensionless kernel $K_{\rm ML}$ is given in
equation~(\ref{KML}). 

\subsubsection{Case of Diemand-Moore-Stadel anisotropy}

Finally for the anisotropy profile (eq.~[\ref{betaDMS}])
proposed by \cite{DMS04_vel}, 
equation~(\ref{genMofr}) leads to (with $D_\beta$ from Table~\ref{shortcuts})
\begin{equation}
- \left ({a^{1/3}-r^{1/3}\over a^{1/3} }\right )
\rho\,{G M\over r^2} (r)
= p'_{\rm iso} - {2/3\over \left (a\,r^2\right)^{1/3}}\,
\left ({5\,a^{1/3} - 3\,r^{1/3}\over a^{1/3} - r^{1/3}}\right )
\left [p_{\rm iso} + {1/\pi\over \left (a^{1/3}-r^{1/3}\right)^3}\,
\int_r^a
{{\rm d}P\over {\rm d}R}\,K_{\rm {DMS}} \left ({R\over a},{a\over a} \right
    )\,{\rm d}R \right ] \ ,
\end{equation}
for $r<a$, and to the radial solution (eq.~[\ref{vcrad2}]) for $r >a$. 





\subsubsection{General form of the mass profile for specific anisotropy profiles}

Inserting equations~(\ref{pprimeiso}) and (\ref{qgenspec}) into
equation~(\ref{genMofr}), one can obtain a general form for the mass
profiles for the constant anisotropy, Osipkov-Merritt, Mamon-{\L}okas, and
Diemand-Moore-Stadel anisotropy profiles:
\begin{equation}
 -[1-\beta(r)]\,\rho(r)  {G M(r) \over r^2} 
=
        {1\over\pi\,r}\,\int_r^\infty \left \{
{D_\beta(r)\over r}
\,\left [{C_\beta(r)}\,K_\beta-{r\over \sqrt{R^2-r^2}} \right ]\,
{{\rm d}P\over {\rm d}R} - {R\over \sqrt{R^2-r^2}}\,
{{\rm d}^2P\over {\rm d}R^2} \right \}\,{\rm d}R \ ,
\label{Mgenspec}
\end{equation}
where, for the Diemand-Moore-Stadel anisotropy profile, the anisotropy radius
$a$ should be used for the upper integration
limits.
Equation~(\ref{Mgenspec}) allows to express the mass profile as a unique
single integral of the observations, where $C_\beta(r)$ and 
$D_\beta(r)$
are given in Table~\ref{shortcuts}, while the kernel $K_\beta$ is given by
equations~(\ref{Kcst}), (\ref{KOM}), (\ref{KML}), and (\ref{KDMS}) for the
constant anisotropy,  Osipkov-Merritt, Mamon-{\L}okas, and
Diemand-Moore-Stadel anisotropy models, respectively.
Equivalently,
equation~(\ref{Mgenspec}) can be used to formulate the circular velocity
profile
\begin{equation}
v_c^2(r) = {1\over \pi\,\left [1-\beta(r)\right]\,\rho(r)}
\,\int_r^\infty \left \{
{R\over \sqrt{R^2-r^2}} 
{{\rm d}^2 P \over
    {\rm d}R^2} 
- {D_\beta(r)\over r}\,\left [{C_\beta(r)}\,K_\beta(R,r)-{r\over \sqrt{R^2-r^2}}\right]\,
{{\rm d} P \over
    {\rm d}R} \right \}\,{\rm d}R \ .
\label{vcgenspec}
\end{equation}
For isotropic models, equation~(\ref{vcgenspec}) with $D_\beta = C_\beta =
K_\beta = 0$ recovers the second equality in equation~(\ref{vciso}).

In practice, writing the tracer density as
$\rho(r) = \rho(a)
\,\widetilde \rho(r/r_s)$, 
where $r_s$ is the characteristic scale of the tracer,
the projected pressure as $P(R) = P(r_s)\,\widetilde P (R/r_s)$,
equation~(\ref{vcgenspec}) yields
\begin{eqnarray}
\left [{v_c(r)\over \sigma_{\rm los}(r_s)}\right]^2
&\!\!\!\!\!\!=\!\!\!\!& {\Sigma(r_s)/\left[\pi\,r_s\,\rho(r_s)\right]\over
\left [1-\beta(r)\right]\,\widetilde\rho(r/r_s)}
\,\int_x^\infty \left \{
{X\over \sqrt{X^2-x^2}} 
{{\rm d}^2\widetilde P \over
    {\rm d}X^2} 
- {D_\beta(r_s x)\over x}\,\left [{C_\beta(r_s x)}
\,K_\beta-{x\over \sqrt{X^2-x^2}}\right]\, 
{{\rm d}\widetilde P \over
    {\rm d}X} \right \}\,{\rm d}X \ ,
\nonumber \\
&\!\!\!\!\!\!=\!\!\!\!& {\Sigma(r_s)\over \left[\pi\,r_s\,\rho(r_s)\right]}
\,{r/r_s\over
\left [1-\beta(r)\right]\,\widetilde\rho(r/r_s)}
\nonumber \\
&\mbox{}& \times
\int_0^{\cosh^{-1} \left({X_{\rm max}/ x}\right)}
\!\!\!\!\widetilde P''(x\,\cosh u)\,\cosh u - {D_\beta(r_s x)\over x}\,
\left [{C_\beta(r_s x)}\,K_\beta(x \cosh u,x)\,\sinh u-1\right
]\,\widetilde P'(x \cosh u)\,du
\label{vcgenspecnorm} 
\end{eqnarray}
where $x = r/r_s$, $X = R/r_s$, and where the second
equality of equation~(\ref{vcgenspecnorm}) is useful to avoid the singularity
at $X=x$, integrating out to the equivalent of, say, $10\,r_v$, i.e. $X_{\rm
  max}= 10\,r_v/r_s$.
All quantities on the right-hand side of the two equalities in
equation~(\ref{vcgenspecnorm}) are
known or assumed (the anisotropy profile). In particular, the numerator of the
factor in front of the integral of the first equality of
equation~(\ref{vcgenspecnorm}) is a function of the shape of the tracer 
density profile, found by Abel inversion (eq.~[\ref{rhoder}]) of the surface
density profile.

\section{Tests}
\label{tests}

\subsection{Accuracy}
We test our mass inversion equations, on
four anisotropy models: isotropic, constant, Osipkov-Merritt and
Mamon-{\L}okas.
For each of these anisotropy models, we compute the projected pressure using
equation~(\ref{Pwkernel}), with the kernels given by \cite{ML05b,ML06b},
evaluated
on a logarithmic
grid from $r = 0.01\,r_s$ to $100\,r_s$ in steps of 0.2 dex.
The projected pressures $P(R)$ were differentiated after cubic spline
interpolation and the integral of equation~(\ref{vcgenspec}) was performed in
steps of $\cosh^{-1} (R/r)$ out to $100\,r_s$. 
We choose our mass and  anisotropy models by placing ourselves in the context of
clusters of galaxies. 
We assume  a one-component NFW model \citep{NFW96}, for which the dimensionless
density and mass profiles can be expressed as
\begin{eqnarray}
\widetilde \rho(x) &=& {\rho(x r_s)\over M(r_s)/\left(4\pi r_s^3\right)} =
           {\left (\ln 2 - 1/2\right)^{-1}\over
  x (x+1)^2} \ ,\\
\widetilde M(x) &=& {M(x r_s)\over M(r_s)} = {\ln(x+1)-x/(x+1)\over \ln
  2-1/2} \ ,
\label{MNFW}
\end{eqnarray}
where $r_s$ is the scale radius, where the slope of the density profile is
$-2$.
We make no use of our assumption that the total and tracer density profiles
are proportional.

The anisotropy profile for dark matter particles in $\Lambda$CDM halos of the
masses of clusters is close to the
Mamon-{\L}okas model \citep{ML05b,Wojtak+08,MBM09} with anisotropy radius
$a\simeq 0.18\,r_{200}$ (\citeauthor{ML05b}) or $0.275\,r_{200} = 1.1\,r_s$
(\citeauthor{MBM09}). We adopt a scaling of $a = r_s$ for the Mamon-{\L}okas
model and for the Osipkov-Merritt model as well, and we adopt a constant
anisotropy model that is fairly radial but consistent with the anisotropy
seen in $\Lambda$CDM halos: $\beta = 0.4$.

\begin{figure}
\centering
\includegraphics[width=0.49\hsize]{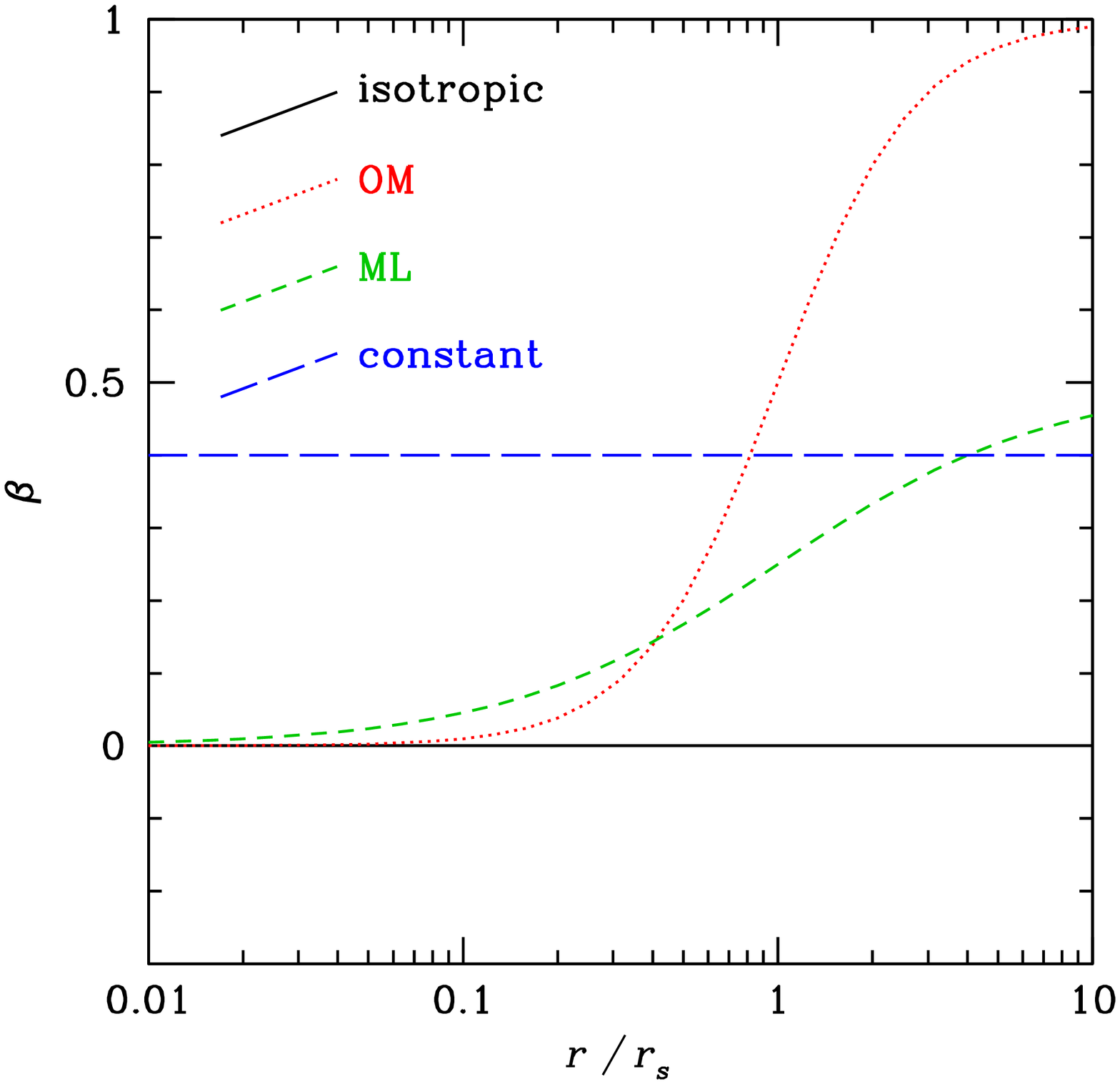}
\includegraphics[width=0.49\hsize]{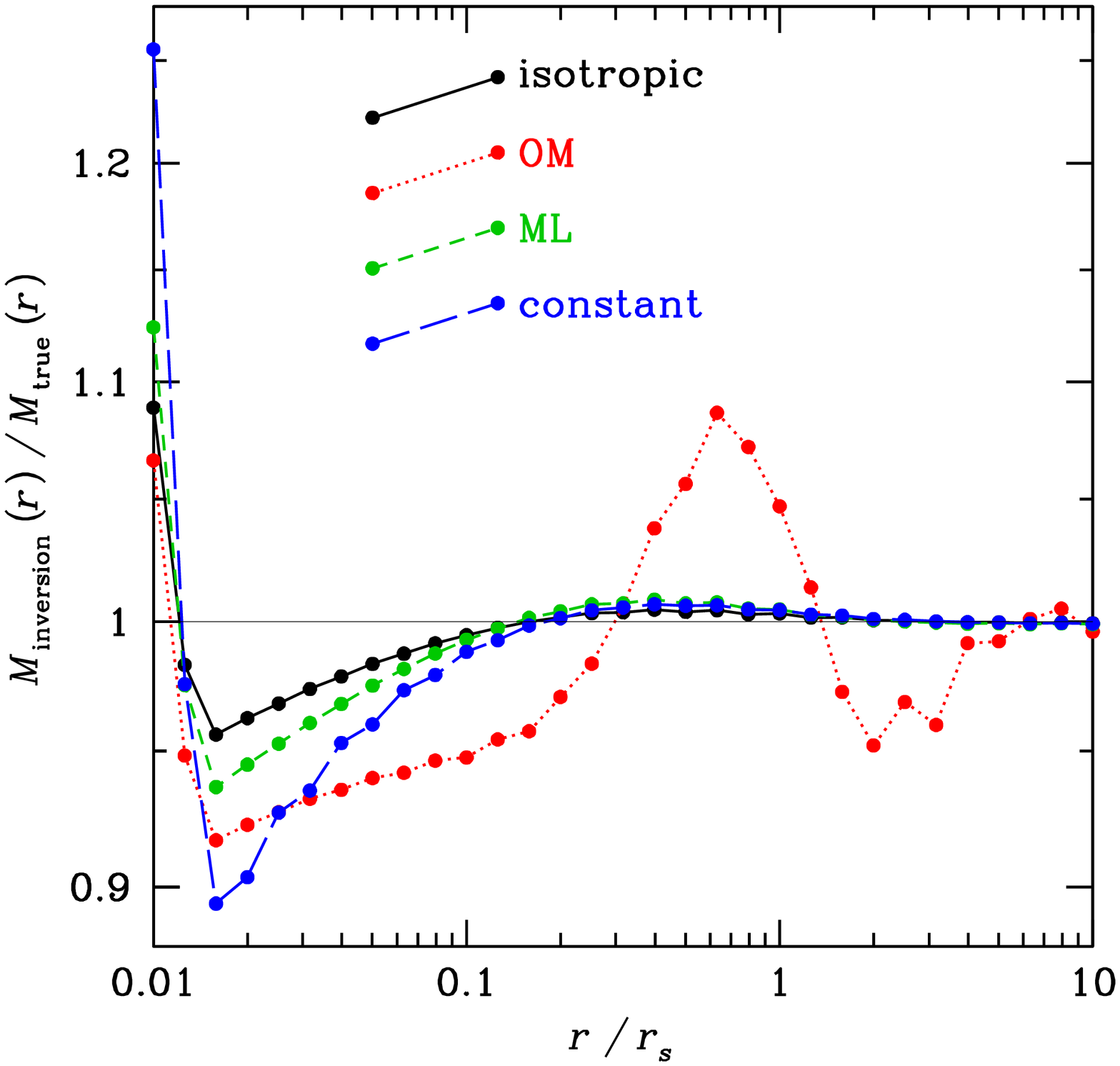}
\caption{\emph{Left:} Adopted anisotropy profiles:
isotropic (\emph{solid black line}),
Osipkov-Merritt (eq.~[\ref{betaOM}], with $a = r_s$, \emph{dotted red line}), 
Mamon-{\L}okas (eq.~[\ref{betaML}], with $a = r_s$, \emph{short dashed green line}),
and
$\beta = \rm cst = 0.4$ (\emph{long dashed blue line}).
\emph{Right:} Accuracy of the mass inversion (in the absence of noise):
ratio of inferred (eq.~[\ref{vcgenspec}], using eq.~[\ref{Pwkernel}]
    to first evaluate $P$ on logarithmic grid of 0.2 dex steps,
    and using the dimensionless functions of
    Table~\ref{shortcuts}, and the dimensionless kernels of
    equations~[\ref{KOM}], [\ref{KML}], and [\ref{Kcst}], for the latter three
    anisotropy models) 
over true NFW mass profiles for the four
  anisotropy models shown in the left panel.
\label{mratfig}}
\end{figure}

Figure~\ref{mratfig} shows the comparison of the circular velocity profiles
obtained from 
the mass inversion equation~(\ref{vcgenspec})  with the true circular
velocity profiles. 
Despite the double differentiation of the projected pressure, the circular
velocity (hence mass) profiles are recovered to a few percent relative
accuracy or better,\footnote{Both mass
  inversion and deprojection appear to be unstable at radii $r<r_s$
for the Mamon-{\L}okas
  anisotropy model when $a$ is exactly set to $r_s$, when using our
  {\tt Mathematica} routines (but this odd behavior is not present when tested with
  other software). The figure shows the
  case $a = 1.001 \,r_s$.} except at the innermost point where the mass is
overestimated by 4 to 10\% in the four anisotropy models, because of the
inaccurate cubic-spline interpolation of $P(R)$ near the edges.
The accuracy of the mass inversion is even better if we use
a finer grid to measure the projected pressure 
before the cubic spline interpolation of
$P(R)$ and subsequent mass inversion: for
example with the OM anisotropy, the maximum relative error in the recovered
mass decreases with grid
size from 9\% (0.2 dex steps) to 0.6\% (0.02 dex steps).

\subsection{Robustness to small data samples}

We next test the accuracy of the recovered mass profiles when the data is
sparse and noisy.
We consider the case of velocity measurements in a cluster of galaxies. We
assume that the cluster has 500 measured velocities within $5\,r_s$ (which is
roughly the cluster virial radius), and assume for simplicity that we have
line-of-sight velocity dispersions measured in 10 equal size radial bins 
centered from 0.25 to $4.75 \,r_s$.
With $N=50$ velocities per bin, 
the velocity
dispersions are known to a relative accuracy of $\sqrt{1/2/( N\!-\!1)} = 10.1\%$
(e.g. \citealp{Lupton93}), 
and we fold this noise\footnote{We neglect the noise on the uncertain surface
  density profile, which contributes negligibly to the noise in the projected
  pressure in comparison to the noise in
  the velocity dispersion.}
into
the predicted line-of-sight velocity dispersion profile, using the same
seed for the random number generator for all four anisotropy profiles.
We extrapolate the projected pressures to larger radii 
by fitting a power-law to $P(R)$
using the last 5 data points, at outer linearly spaced outer radii, with the
same spacing as the data, and then fit a 4th order polynomial to the set of
observed and mock-extrapolated data. 
We repeated these tests 5 times with different seeds for the random generator.

\begin{figure}
\centering
\includegraphics[width=0.49\hsize]{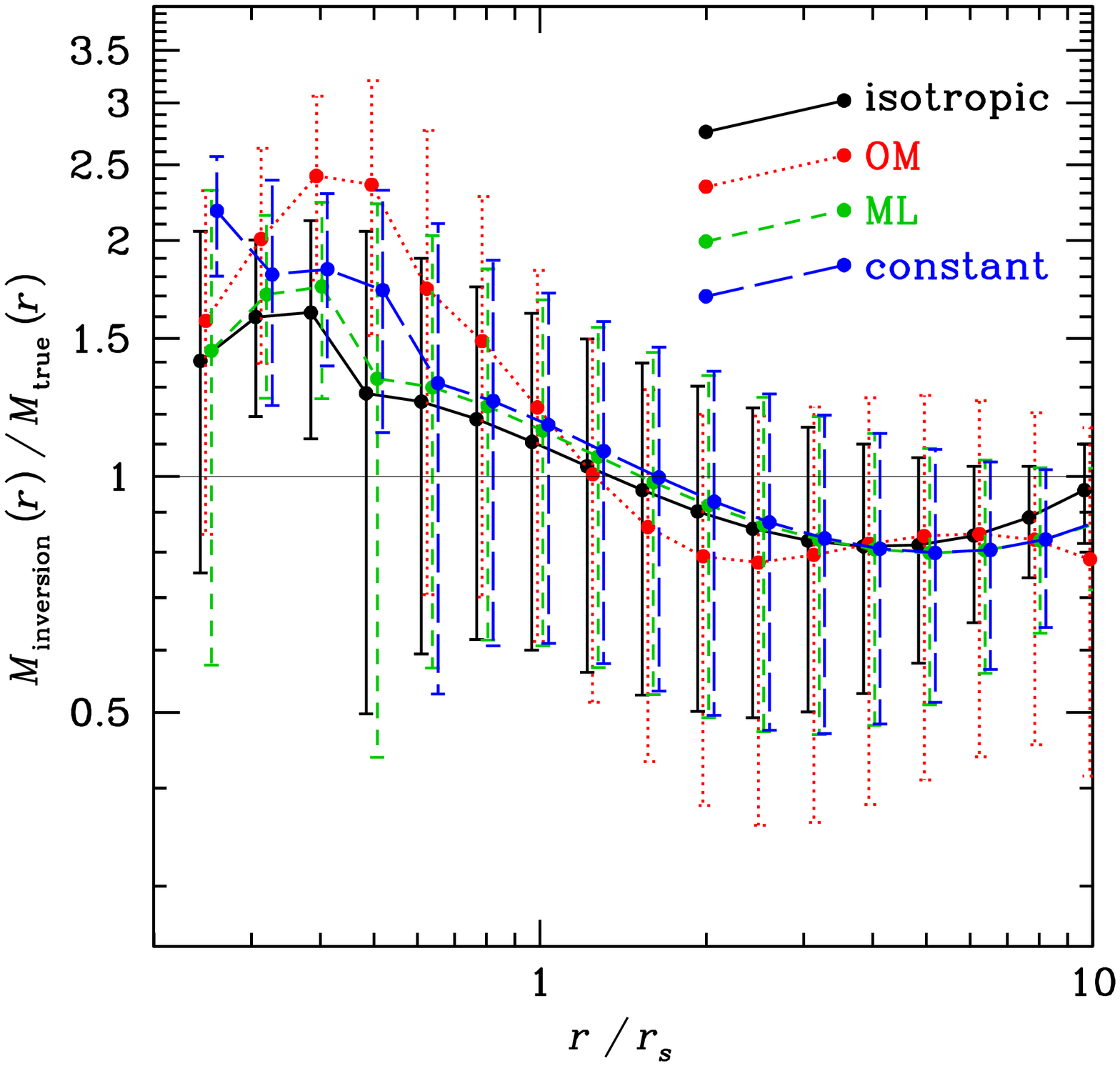}
\includegraphics[width=0.49\hsize]{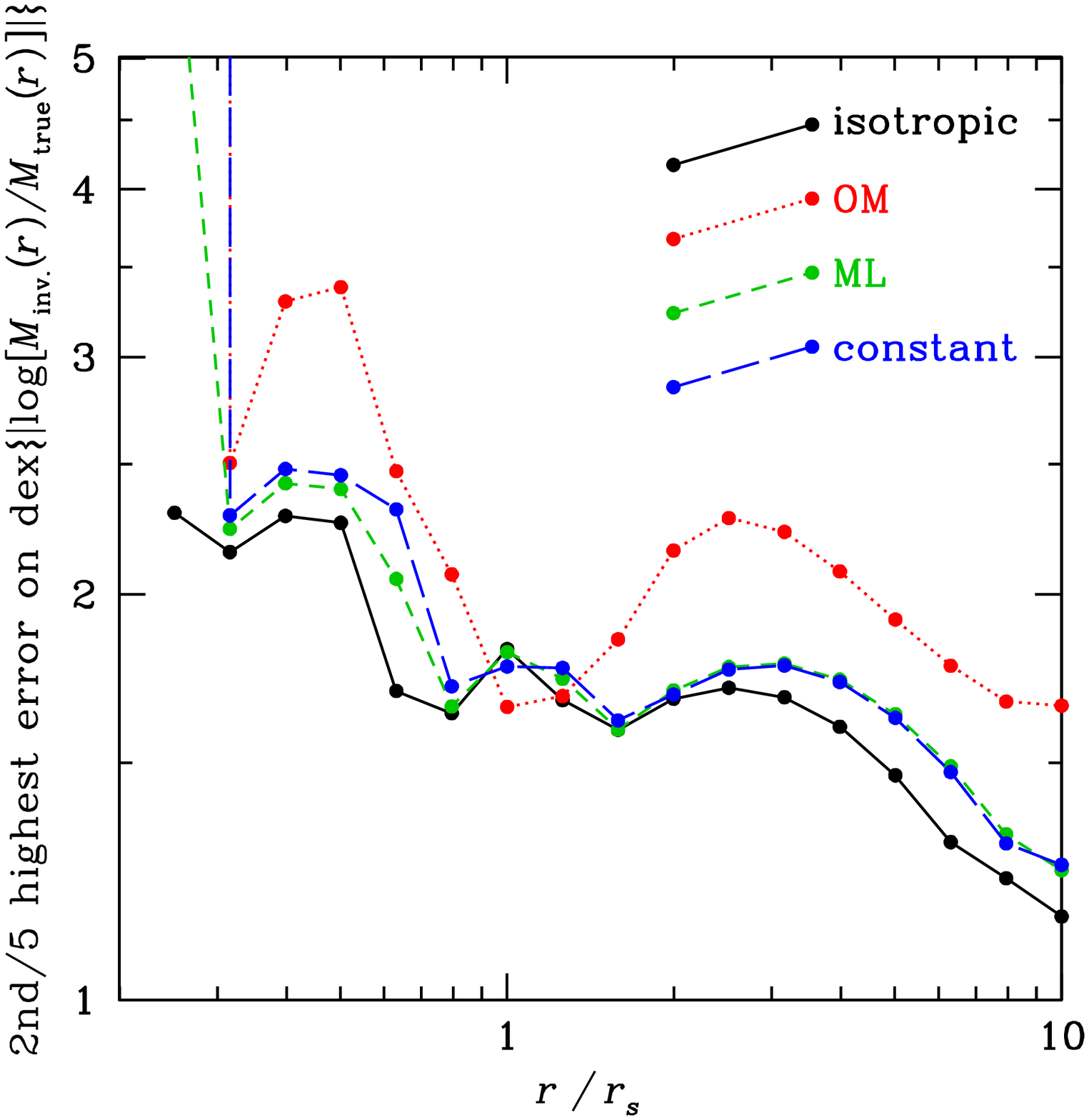}
\caption{Robustness of the mass inversion to small data samples.
\emph{Left:} Same as right panel of Fig.~\ref{mratfig}, 
but for the projected
  pressure profile measured on a linear grid of 10 radial bins 
from 0.25 to $4.75\,r_s$, with 20\% relative
  gaussian
errors on the projected pressure
(i.e. 10\% errors on the line-of-sight velocity dispersion measurements  based
upon 50 velocities per radial bin).
The error bars show the standard deviations on 5 tests with different seeds
for the random number generator.
The points and error bars are slightly shifted along the $x$-axis for clarity.
\emph{Right:} 2nd highest error out of 5 tests on recovered mass profile.
A value of unity indicates a perfect recovery of the mass.
\label{mratclusnoise}}
\end{figure}

The left-hand panel of 
Figure~\ref{mratclusnoise} shows the accuracy of the mass inversion is much
worse than in the academic case with no noise.
In particular, the extrapolation errors at radii lower than the lowest data
point make the inner mass profile inaccurate to factors often greater than
2. With the isotropic, $\beta=0.4$, and Mamon-{\L}okas anisotropy models, the
mass profile is nevertheless recovered to typically better than 20\% accuracy
for $r > 0.8\,r_s$, out to twice the radius of the last data point.
However, the large error bars show that there is a large scatter
in the accuracy of the recovered mass profile for different randomly
generated projected pressure profiles. 
The right-hand panel of 
Figure~\ref{mratclusnoise} gives the second highest error among the five tests
performed, for each given radius and anisotropy model. Typical such
80-percentile errors are of the order of 70\% for $r > 0.8\,r_s$.
Surprisingly, this typical error decreases to only 20\% at high radii ($r >
8\,r_s$), despite the fact that the projected pressure is extrapolated beyond
$r = 4.5\,r_s$.

\subsection{Robustness to the wrong anisotropy model}

The essential ingredient to the mass inversion is the knowledge of the
velocity anisotropy profile. How wrong can the mass inversion be if the
incorrect anisotropy profile is used?
We adopt the Mamon-{\L}okas anisotropy model with $a=r_s$ similar to what is
found for cluster-mass $\Lambda$CDM halos \citep{MBM09} and compute the
projected pressure for an NFW model with this anisotropy model.
We then
perform the mass 
inversion assuming other anisotropy profiles to see how off we are. In this
exercise, we assume perfect data, i.e. no noise.

\begin{figure}
\centering
\includegraphics[width=0.49\hsize]{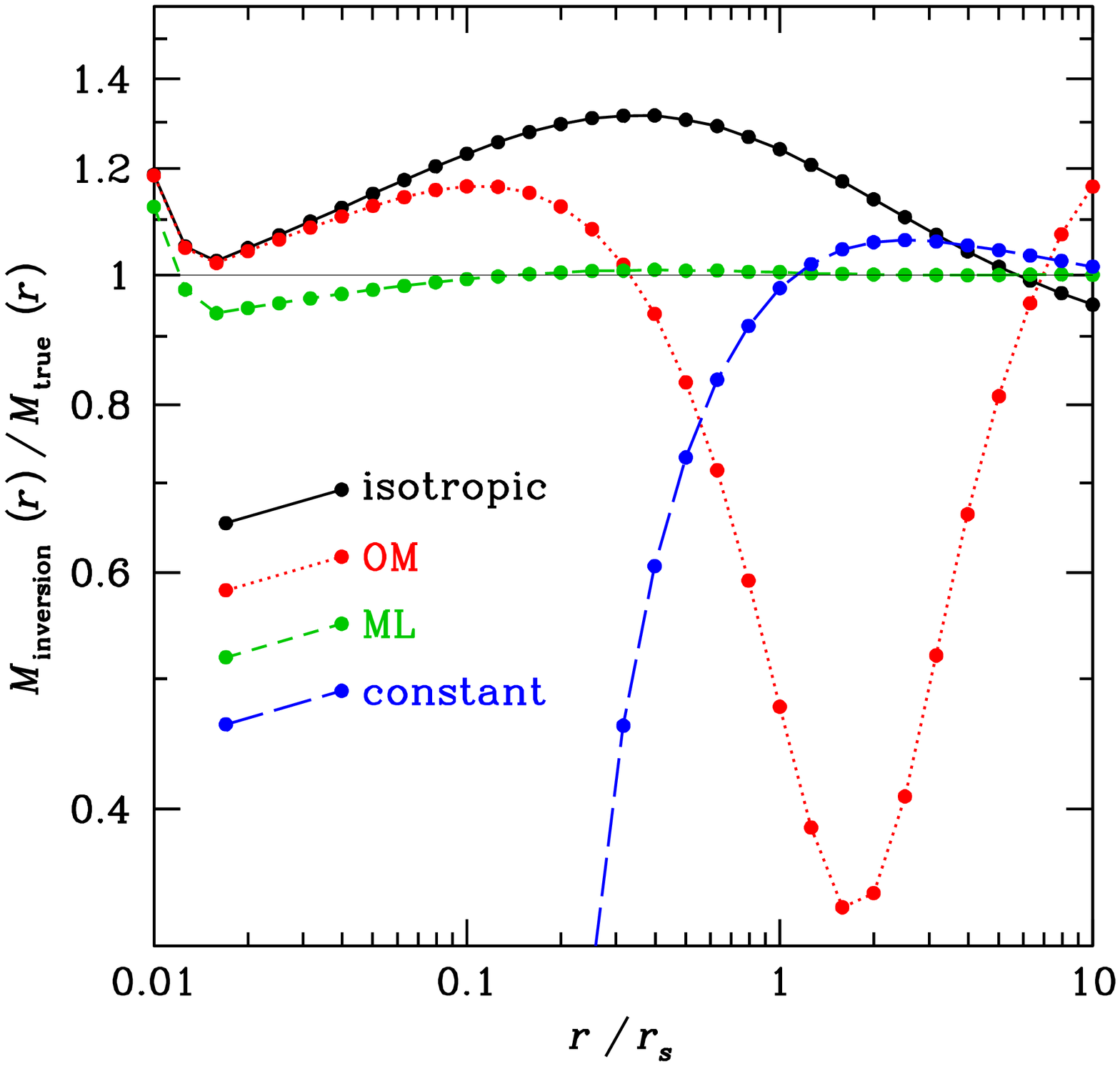}
\includegraphics[width=0.49\hsize]{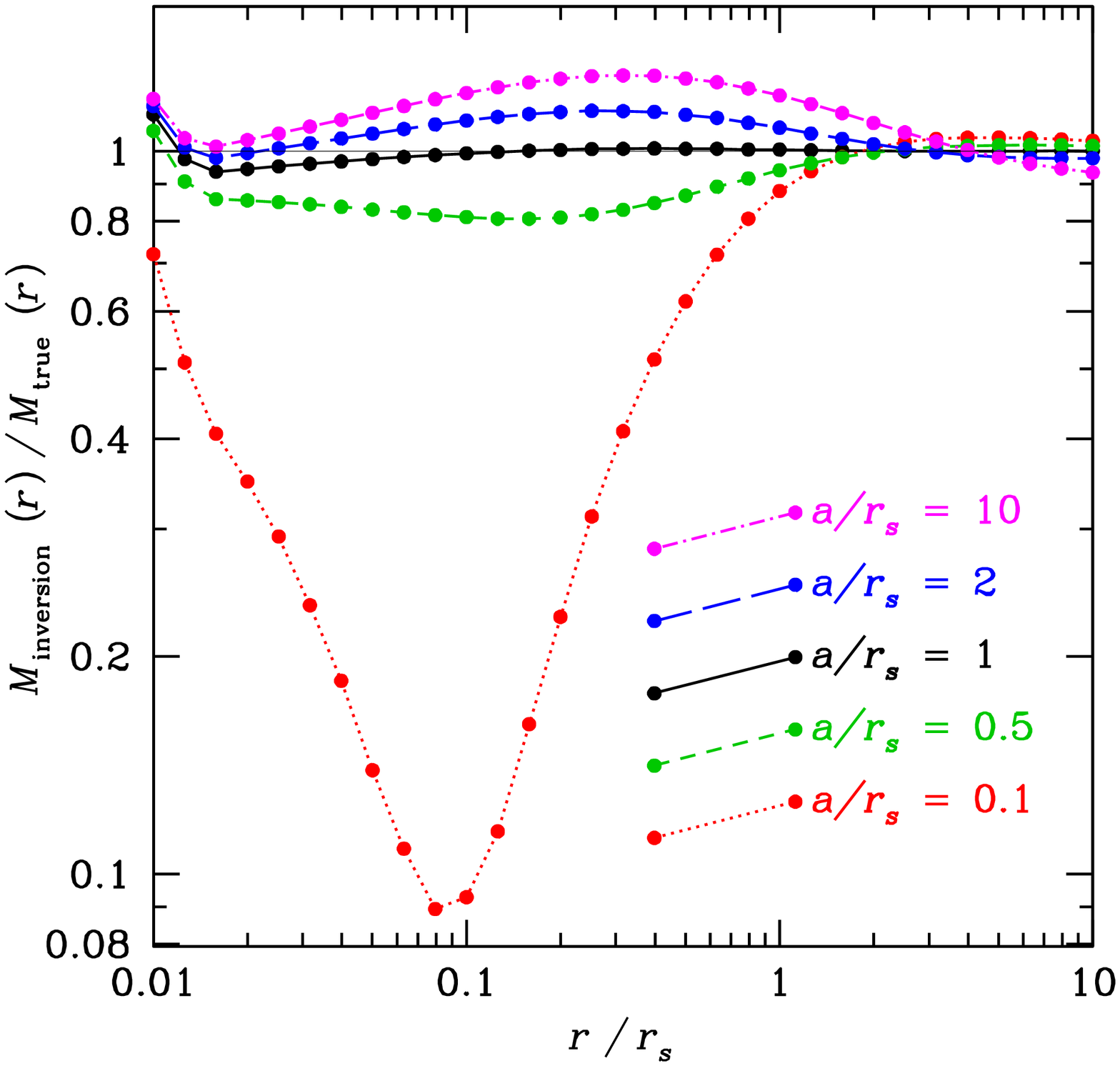}
\caption{Robustness of the mass inversion to the wrong choice of anisotropy profile.
\emph{Left:} Same as right panel of 
Fig.~\ref{mratfig}, where the true anisotropy profile is now always an
$a=r_s$ Mamon-{\L}okas model, but assuming that of the other three anisotropy
models.
\emph{Right:} Same as left panel, but where the assumed anisotropy profile is
always the $a=r_s$ 
Mamon-{\L}okas model, but with 5 different choices for $a/r_s$.
\label{wronganis}}
\end{figure}

The left panel of 
Figure~\ref{wronganis} shows that the mass profile is recovered to better
than 33\% accuracy for all anisotropy models at $r>4\,r_s$, i.e. beyond the
virial radius. Within the virial radius, the Osipkov-Merritt 
underestimates the mass by as much as a factor 3 around $2\,r_s$, but is much
more precise at very low radii.
The $\beta=0.4$ model is accurate for $r>r_s$, as expected as it resembles
there the Mamon-{\L}okas model, but underestimates the mass by increasingly
large factors at radii $r<r_s$, and the recovered mass actually goes negative
at $r<0.17\,r_s$.
Finally, the isotropic model finds the correct mass to within 30\% at all
radii, usually overestimating the true mass.
Interestingly, at $r\simeq 7\,r_s$, all four anisotropy models lead to the
correct mass to within 5\%. 

The right panel of Figure~\ref{wronganis} indicates that the recovered mass
is not very sensitive to the assumed anisotropy radius, as the mass is
recovered to 20\% accuracy, unless the anisotropy radius is assumed to be 10
times lower than it actually is.
This graph also shows that at $r
\simeq 3\,r_s$ (i.e. roughly two-thirds the virial radius of clusters), the
mass is correctly recovered to better than 5\% for our five choices of
anisotropy radius.

\section{Discussion}

The mass inversion algorithm presented in this work generally (eq.~[\ref{M2}]), 
requires two steps: 1) deprojection and 2) inserting the
radial 
pressure in the Jeans equation to derive the mass. The deprojection
(eq.~[\ref{qgen4}]) requires a single integral involving of a quantity,
$p_{\rm iso}$ 
(eq.~[\ref{pisodef}]) that is itself a single integral involving the derivative
of the observed projected pressure.
The second step (mass inversion) also
requires a single integral involving the derivative of $p_{\rm iso}$.
Each differentiation of the data introduces errors, and the full mass
inversion requires three single integrals.
For the special cases of
simple anisotropy models, we find it preferable to write the mass profile
with a single
integral involving the double derivative of the observed projected
pressure. Indeed, this requires a single smoothing operation before
differentiation, thus leading to more accurate results, even if the
mathematical formulation of the deprojections and mass inversions for each of
the simple anisotropy models 
has strongly increased the number of equations in this
article. 

While this work (and \cite{Wolf+09}) used the Abel inversion for the
kinematic deprojection, one 
can alternatively apply Fourier methods (see also \citealp{KN88} and Kalnajs
cited in \citealp{SBM96}).
Indeed, structural and kinematic
  projection can be written as a convolution:
\[
F(X) = \int_{-\infty}^{\infty} f(x)
\,K(X-x)\,{\rm d}x
\ ,
\]
where $X=R^2$, $x=r^2$, $F(X)$ and $f(x)$ correspond to either $\Sigma(R)$
and $\rho(r)$ (structural
projection) or $P(R)$ and $(1-\beta)\,p + \int_r^\infty \beta\, p\, {{\rm d}s/
  s}$ (anisotropic 
kinematic projection, following \citeauthor{Wolf+09}, which simplifies to
$p$ for isotropic kinematic projection), and where
\[
K(y) = \left \{ \begin{array}{ll}
\displaystyle (-y)^{-1/2} & y \leq 0 \ , \\
\displaystyle 0 & y > 0 \ .\\
\end{array}
\right .
\]
Hence, with the convolution theorem, deprojection is obtained by applying an
inverse Fourier transform to
\[
\widetilde f(\omega) = {1\over
  \sqrt{2 \pi}}\,{\widetilde F(\omega)\over \widetilde K(\omega)} = (1+i)\,\hbox{sgn}(\omega)\,
\sqrt{\omega\over 2\pi}\,
\widetilde F(\omega) 
= [1+i\,\hbox{sgn}(\omega)]\,\sqrt{|\omega|\over 2\,\pi}\,\widetilde F(\omega)\ ,
\]
where $\widetilde f(\omega)$, $\widetilde K(\omega)$ and $\widetilde
F(\omega)$ are the Fourier transforms of $f(x)$, $K(y)$ and $F\left(|X|\right)$,
respectively (note the absolute values in the last term).
A comparison of the accuracy of the two deprojection 
techniques is beyond the scope of the
present article.

Our mass inversion algorithm should serve as a useful
technique to get around the mass-anisotropy degeneracy in the case where the
anisotropy profile is thought to be known.
As mentioned in the end of Sect.~I, there is a good convergence on the
anisotropy profiles of $\Lambda$CDM halos as well as those of elliptical
galaxies formed by binary mergers of spiral galaxies.
Moreover, the anisotropy profile in many simulations appears linearly related
to the slope of the (tracer) density profile \citep{HM06}, and this can be
used to lift the mass-anisotropy degeneracy.
A first application of our algorithm was given by \cite{BS06} for the
analysis of stacked clusters of galaxies.

The mass inversion technique has the advantage of producing a
non-parametric\footnote{Although the algorithm uses a parametrized anisotropy profile and a
  smooth fit through the projected pressure profile, the mass profile
  that comes out is non-parametric.}
mass profile, which can then be used to test the popular parametrizations of
the mass profile (or alternatively of the density profile, the circular
velocity profile or the density-slope profile).

In Sect.~\ref{tests}, we show that, for a mock NFW galaxy cluster with mildly
increasing radial velocity anisotropy as seen in $\Lambda$CDM halos and with
typical line-of-sight velocity
dispersion profiles, measured with 50 velocities per radial bin, the mass
inversion should be accurate to typically better than 70\% relative errors at
most radii and better than 20\% for anisotropy models other than the
Osipkov-Merritt one at $r > 8\,r_s$.
The relatively high errors are a consequence of the double derivative of the
observed projected pressure, ${\rm d}^2P/{\rm d}R^2$, that enters the mass
inversion equation~(\ref{M1}) or (\ref{M2}), through the term $p_{\rm iso}$,
or in equation~(\ref{Mgenspec}) or (\ref{vcgenspec}) for the special cases of
anisotropy models.
The errors are high at radii smaller than the first radial bin of the
observed line-of-sight velocity dispersion profile. This illustrates the
concept that kinematical modelling can only recover the mass and anisotropy
at radii corresponding to the projected radii of the data. Nevertheless, with
power-law extrapolations of the data to outer radii, we show that the mass
inversion can recover mass profiles with good accuracy far 
beyond the outermost data point.
Note that the mass inversion involves integrals out to infinity
(e.g. eq.~[\ref{vcgenspec}]), so one expects that the method should be most accurate when
the tracer density profile falls fast at large radii. 
Our use of the NFW model for the tracer, with its shallow
outer slope of ${\rm d}\ln\rho/{\rm d}\ln r=-3$ is
thus 
expected to provide poorer results for the mass inversion than for steeper
tracer density profiles.

We found that the recovered mass is correctly returned, 
independently of the shape of the anisotropy profile at $r=7\,r_s$, 
and independently of the anisotropy
radius for our chosen anisotropy model at $r = 3\,r_s$.
A similar independence of the recovered mass on
the assumed anisotropy profile 
has been recently noticed by \cite{Wolf+09}
in the context of dwarf spheroidal and elliptical 
galaxies (for which the dark matter may
not follow the stars, which themselves do not follow the NFW model).
However, \citeauthor{Wolf+09} prove analytically that
this robustness to the anisotropy model occurs near the 
the radius of slope $-3$. Now, the NFW model has shallower slopes everywhere,
reaching $-3$ at infinite radius.
\citeauthor{Wolf+09} notice that, for density profiles similar to those of
ellipticals and dwarf spheroidals, the radius of slope $-3$ is close
to the half-mass radius.
In contrast, 
in the current context of clusters, the NFW model is divergent in mass
(eq.~[\ref{MNFW}]), and the concept of half-mass radius is
ill-defined. Moreover, 
the radius where the mass is recovered for all anisotropy models tested
is at 7 scale radii, which is outside the virial radius, hence not comparable
to the half-light radius of elliptical and dwarf spheroidal galaxies. Fixing
the anisotropy to the 
Mamon-{\L}okas model (which \citealp{ML05b} found to be a good fit to the
anisotropy profile of the halos in $\Lambda$CDM cosmological simulations),
the recovered mass is most robust to the anisotropy 
radius at $3\,r_s$, which is roughly two-thirds of the cluster virial radius,
again not directly comparable to the half-light radius of dwarf spheroidals
and ellipticals.

The mass inversion technique is thus a useful complement to the set of tools
one has to lift the mass-anisotropy degeneracy in spherical systems. Mass
inversion  is
certainly not the privileged tool, but should be considered as one of many
tools for the \emph{exploratory data analysis} of spherical systems viewed in
projection, in addition to anisotropy inversion, fitting models to the
line-of-sight velocity dispersion and possibly kurtosis profiles, and fitting
models, distribution functions, orbits and $N$-body systems to the
distribution of particles in projected phase space. Ideally, one would analyze
the kinematics of spherical systems using a variety of these tools. We are
preparing such global analyses on dwarf spheroidal and elliptical galaxies,
as well as on groups and clusters of galaxies.

\section*{Acknowledgments}

We thank Andrea Biviano for suggesting one of us (G.A.M.) to
estimate $\rho\,\sigma_r^2$ given $\Sigma\,\sigma_{\rm los}^2$, i.e. to
perform the kinematical deprojection, Joe Wolf for useful discussions
during the final stage of this work (prompting us to return to this work
after much neglect for over a year), as well as Aaron Romanowsky for useful
comments. 
We are highly indebted to 
Richard Trilling for his critical reading of the
manuscript, which helped us spot and fix several mathematical errors, and to
the referee, Prasenjit Saha, for pointing out to us that projection is
essentially a convolution and for his comments
that improved the readability of the manuscript.
\bibliography{master}

\appendix
\section{Abel deprojection}
\label{deprojisoapp}
In this appendix, we remind the reader of the derivation of the deprojection
of equation~(\ref{abelproj})
with the Abel inversion.
Consider
\begin{equation}
J(r) = \int_r^\infty {\Sigma(R) \,R\, {\rm d}R \over (R^2-r^2)^{1/2}} \ .
\label{Jofsorig}
\end{equation}
Replacing $\Sigma(R)$ in
equation~(\ref{Jofsorig}) by its definition in
equation~(\ref{abelproj}), one finds, after inverting the order of integration:
\begin{equation}
J(r) = 
2\,\int_r^\infty \rho(s) \,s\,{\rm d}s \,\int_r^s {R\,{\rm d}R\over \left (R^2-r^2\right
)^{1/2} \,\left (s^2-R^2\right )^{1/2}} \ .
\label{Jdes}
\end{equation}
The internal 
integral in equation~(\ref{Jdes}) is equal to
$\pi/2$, as inferred from the substitution $\sin^2 \theta = (R^2-r^2)/(s^2-r^2)$.
Hence,
\[
J(r) = \pi \,\int_r^\infty \rho(s)\,s\,{\rm d}s \ ,
\]
and therefore
\begin{equation}
\rho(r) = - {1\over \pi r} {{\rm d}J\over {\rm d}r} \ .
\label{rhor0}
\end{equation}

Now integrating equation~(\ref{Jofsorig}) by parts, one gets
\begin{equation}
J(r) = \lim_{R\to\infty} \sqrt{R^2-r^2}\,\Sigma(R) - \int_r^\infty {{\rm d}\Sigma\over {\rm
    d}R}\,\left 
(R^2-r^2\right)^{1/2}\,{\rm d}R \ ,
\label{Jofsparts}
\end{equation}
For all realistic density profiles, $\Sigma(R)$
falls faster than $R^{-1}$, as is the case for the surface density profiles
of globular clusters, elliptical galaxies and clusters of galaxies.
Hence, the surface term in equation~(\ref{Jofsparts}) is zero and one can
write 
\begin{equation}
{{\rm d}J\over {\rm d}r} = r\,\int_r^\infty  {{\rm d}\Sigma\over {\rm
    d}R}\,{{\rm d}R\over \left (R^2-r^2\right )^{1/2}} \ .
\label{dJdr}
\end{equation}
%
%
Inserting the derivative of $J$ of equation~(\ref{dJdr}) into
equation~(\ref{rhor0})  
leads to equation~(\ref{rhoder}).
The surface term that survived when $\Sigma \propto 1/R$ disappears in the
derivative. 

\section{Kinematic deprojection for the tangential dynamical pressure}
\label{appq}

In this appendix, we derive equations~(\ref{qgen3}) and (\ref{qgen4}) for the
tangential dynamical pressure.

Differentiating equation~(\ref{impliciteqp}), one finds
to get the differential equation
\begin{equation}
p_\theta' - {\beta\over 1-\beta}\,{p_\theta\over r}  = p_{\rm iso}' \ .
\label{qprime}
\end{equation}
Now, if we write
\begin{equation}
p_\theta' - {\beta\over 1-\beta}\,{p_\theta\over r} = {1\over f}\,{{\rm d}
(fp_\theta)\over {\rm d}r} \ ,
\label{qprimewf}
\end{equation}
then equations~(\ref{qprime}) and (\ref{qprimewf}) lead to
\begin{equation}
p_\theta(r) = -{1\over C_\beta(r)}\,\int_r^\infty f\,p'_{\rm iso}\,{\rm d}s
\ ,
\label{q1}
\end{equation}
where the upper limit at infinity ensures that $p_\theta = (1-\beta)\,\rho\,\sigma_r^2$
does not reach negative values at a finite radial distance.
But equation~(\ref{qprimewf}) directly gives
\[
{{\rm d}\ln f\over {\rm d}\ln r} = - {\beta(r) \over 1-\beta(r)} \ ,
\]
hence
\begin{equation}
g(r) = g(r_1) \,\exp \left (-\int_{r_1}^r {\beta \over
1-\beta}\,{{\rm d}s\over s} \right ) 
\label{fq}
\end{equation}
for any arbitrary $r_1$.
With equation~(\ref{fq}), equation~(\ref{q1}) allows one to recover 
equation~(\ref{qgen3}): 
\begin{eqnarray}
p_\theta(r) 
&=& -\exp\left (\int_{r_1}^r {\beta\over 1-\beta}\,{{\rm d}s\over s} \right )
\,\int_r^\infty \exp \left (-\int_{r_1}^s {\beta\over 1-\beta}\,{{\rm d}t\over t}
\right )\,p'_{\rm iso}\,{\rm d}s \ , \nonumber \\
&=& 
- \int_r^\infty \exp \left (-\int_r^s {\beta\over 1-\beta}\,{{\rm d}t\over t}
\right )\,p'_{\rm iso}\,{\rm d}s \ ,
\label{qgen1}
\end{eqnarray}
where the second equality is obtained adopting $r_1=r$.

Integrating by parts  the integral in
equation~(\ref{qgen1}), we finally recover equation~(\ref{qgen4}) 
\[
p_\theta(r) 
= p_{\rm iso}(r)
- \int_r^\infty p_{\rm iso} {\beta\over
1-\beta}\,
\exp\left(-\int_r^s {\beta\over 1-\beta}\,{{\rm d}t \over t} \right
){{\rm d}s\over s} \ .
\]

\end{document}